\documentclass[
reprint,
aps,
pre,
twocolumn,
superscriptaddress,
showkeys,
showpacs,
]{revtex4-2}
\usepackage{graphicx}
\usepackage{amssymb}
\usepackage{amsmath}
\usepackage{mathtools}
\usepackage{float}
\usepackage[T1]{fontenc}
\usepackage[utf8]{inputenc}
\usepackage[shortlabels]{enumitem}
\usepackage{hyperref}
\usepackage[dvipsnames]{xcolor}
\hypersetup{    % temporarily added by Paul to see links more clearly
    colorlinks,
    linkcolor={blue!80!black},
    citecolor={blue!80!black},
    urlcolor={blue!80!black}
}

\DeclarePairedDelimiter\ceil{\lceil}{\rceil}
\DeclarePairedDelimiter\floor{\lfloor}{\rfloor}

\newcommand{\W}[1]{\mathcal{W}\big[#1\big]}

\newcommand{\uW}[1]{\underline{\mathcal{W}}\big[#1\big]}
\newcommand{\Wvar}[1]{\mathcal{W}_\text{var}\big[#1\big]}

\graphicspath{{./figs/}}


\begin{document}
\title{Weighted averages in population annealing: analysis and general framework}
\author{Paul L. Ebert}
\email{paul.ebert@physik.tu-chemnitz.de}
\affiliation{Institut für Physik, Technische Universität Chemnitz, 09107 Chemnitz, Germany}
\author{Denis Gessert}
\email{gessert@itp.uni-leipzig.de}
\affiliation{Centre for Fluid and Complex Systems, Coventry University, Coventry CV1~5FB, United Kingdom}
\affiliation{Institut f\"{u}r Theoretische Physik, Leipzig University, IPF 231101, 04081 Leipzig, Germany}
\author{Martin Weigel}
\email{martin.weigel@physik.tu-chemnitz.de}
\affiliation{Institut für Physik, Technische Universität Chemnitz, 09107 Chemnitz, Germany}
\date{\today}

\begin{abstract}
    Population annealing is a powerful sequential Monte Carlo algorithm designed to study the equilibrium behavior of general systems in statistical physics through massive parallelism.
    In addition to the remarkable scaling capabilities of the method, it allows for measurements to be enhanced by weighted averaging~\cite{machta:10a}, admitting to reduce both systematic and statistical errors based on independently repeated simulations.
    We give a self-contained introduction to population annealing with weighted averaging, generalize the method to a wide range of observables such as the specific heat and magnetic susceptibility and rigorously prove that the resulting estimators for finite systems are asymptotically unbiased for essentially arbitrary target distributions.
    Numerical results based on more than $10^7$ independent population annealing runs of the two-dimensional Ising ferromagnet and the Edwards-Anderson Ising spin glass are presented in depth.
    In the latter case, we also discuss efficient ways of  measuring spin overlaps in population annealing simulations.
\end{abstract}

\maketitle

%\tableofcontents

\section{Introduction}

    Many successful methodological advances in science are driven by the desire to solve certain notoriously hard problems using contemporary tools.
    In this regard, recent developments in statistical physics with a growing interest in understanding complex systems such as large bio-molecules as well as (structural and spin) glasses are no exception. Meanwhile, powerful supercomputers are clearly among the most notable tools of our time.
    Unfortunately, many of the systems of interest are very hard to simulate, with problems such as divergent relaxation times and the need to average over a large number of disorder samples --- some such systems are even provably NP-hard~\cite{barahona:1982} leaving little hope for exact algorithms applied to problems of reasonable size.
    On the other hand, the limit to which {\em approximate\/} approaches can be pushed is largely decided upon the ability to efficiently use parallel architectures which form the backbone of all modern supercomputers.
    
    Population annealing (PA) can be seen as one of the answers from computational physics to systems displaying frustration, complex free-energy landscapes, slow relaxation times and the associated phase transitions.
    In contrast to more established methods in the field, such as parallel tempering \cite{geyer:91,hukushima:96a}, PA is highly parallelizable and scales almost perfectly~\cite{weigel:17,barash:16,barzegar:18,christiansen:2019} while at the same time providing state-of-the-art algorithmic speed-up for systems with metastability and complex free-energy landscapes~\cite{wang:15a,wang:15b}.
    It was first introduced by Iba and Hukushima~\cite{iba:01,hukushima:03} and later revisited by Machta~\cite{machta:10a}, but closely related  approaches for different applications were also independently proposed in other scientific communities.
    For instance, Zhou and Chen~\cite{zhou:10, zhou:13} introduced an algorithm for continuous global optimization, which is essentially PA applied to a Boltzmann-like distribution.
    In another work that has received significant attention among statisticians, Moral \emph{et al.}~\cite{moral:06} established a very general framework for sequential Monte Carlo (SMC) methods that also includes PA and provides a glimpse into the rich body of related work in mathematical statistics. In molecular physics the technique of diffusion quantum Monte Carlo is an early incarnation of the same idea~\cite{reynolds:82}, only to name a few examples.
    In contrast to the widely adopted Markov chain Monte Carlo (MCMC) approach, PA is sequential in the sense that a \emph{population} of \emph{replicas} (walkers) successively samples from the distributions $\rho_{\beta_0}$, $\rho_{\beta_1}$, $\ldots$ while the control parameter $\beta$ gradually changes into a regime where equilibration is hard.
    The key ingredient to PA is the alternation between an equilibration sub-routine and resampling steps, the latter prioritizing replicas representative of the target distribution at the next value of the control parameter.
    Thus, it is the interplay of preferrential but correlation-inducing resampling and decorrelating but neutral (MCMC) equilibration routines that is responsible and decisive for the efficiency of PA.
    
    In addition to the basic algorithm, a number of improvements for PA have been proposed in recent years~\cite{barash:16,barzegar:18,amey:18,weigel:21}, including adaptive temperature schedules and population sizes among other suggestions. One somewhat underrated aspect is the possibility of combining results from independent PA runs in a way that reduces statistical as well as systematic errors.
    This so-called \emph{weighted averaging} scheme~\cite{machta:10a} thus allows one to obtain better results on limited hardware, to distribute the simulation effort over independent computing systems, and to rectify deficiencies in equilibration {\em a posteriori\/} by adding further runs to the analysis. While this method has been used in studies of spin glasses~\cite{machta:10a,wang:15a}, hard sphere mixtures~\cite{callaham:17,amey:21} and large-$q$ Potts models~\cite{rose:19}, an in-depth analysis of the details of bias reduction and possible pitfalls has not been presented.
    The aim of this work is to close this gap using the Ising ferromagnet and the Ising spin glass in two dimensions as toy models while providing a more comprehensive picture of weighted averaging through new notions, a more explicit notation and enhanced mathematical rigor.
    
    The remainder of this paper is organized as follows. We begin with a detailed introduction to the general PA algorithm in Sec.~\ref{sec:ALG} that should be particularly useful for readers that are new to the topic. Hereafter, the Ising models under consideration and the associated  observables are discussed in Sec.~\ref{sec:MO}, including a comparison of techniques to measure spin overlaps in single PA runs in Sec.~\ref{sec:MO:Overlap}.
    Weighted averaging is introduced in Sec.~\ref{sec:WA}, where we also discuss a general framework for weighted estimators in PA and prove certain asymptotic results.
    Section~\ref{sec:num} starts by summarizing our methodology before various numerical results on the reduction of bias, on intrinsic properties of the weighted-averaging scheme, as well as on statistical errors are reported.
    Finally, a summary of our findings is given in Sec.~\ref{sec:conclusion}.

\section{Algorithm}\label{sec:ALG}

    \subsection{Requirements and basic ideas}\label{sec:ALG:ideas}

        A system to be simulated using PA needs to exhibit a control parameter $\beta$ which determines the equilibrium distribution $\rho_\beta$ on its state space $\Gamma$.
        As $\beta$ is varied throughout the \emph{annealing schedule} $\beta_0$, $\beta_1$, $\ldots$, $\beta_f$, the ratios $\rho_{\beta_i}(\gamma)/\rho_{\beta_{i-1}}(\gamma)$ must (exist and) be known up to a state-independent factor for every $\gamma \in \Gamma$ that is potentially sampled at $\beta_{i-1}$.
        Moreover, one should be able to efficiently sample the initial equilibrium distribution $\rho_{\beta_0}$.
        
        To perform a PA simulation, a \emph{population} of independent states, whose members we call \emph{replicas}, is drawn from $\rho_{\beta_0}$.
        Hereafter, the annealing process begins, i.e., a loop running through a \emph{resampling} step corresponding to $\beta_{i-1} \mapsto \beta_i$, an equilibration algorithm at $\beta_i$, and measurements at $\beta_i$.
        The most crucial part is the resampling which is based on the following observation.
        Suppose that the empirical distribution induced through the population at $\beta_{i-1}$ is $\widehat{\rho}_{i-1}$.
        If $\widehat{\rho}_{i-1} \approx \rho_{\beta_{i-1}}$, a new population close to $\rho_{\beta_i}$ can be created through a copying process by enforcing that the number of copies created of each replica in $\gamma \in \Gamma$ is proportional to $\rho_{\beta_i}(\gamma)/\rho_{\beta_{i-1}}(\gamma)$.
        Thus, if PA is reasonably equilibrated at $\beta_{i-1}$, resampling creates an advantageous initial distribution for equilibration routines at $\beta_i$.
        
        For systems described by the canonical ensemble, as for instance in our numerical work, $\beta$ coincides with the inverse temperature $(k_B T)^{-1}$ and $\rho_\beta$ is the Boltzmann distribution, i.e., $\rho_\beta \propto \exp(-\beta H)$, where $H$ is the Hamiltonian.
        Hence, it is convenient to start the annealing process at infinite temperature $\beta_0=0$, where $\rho_{\beta_0}$ is uniform on $\Gamma$. Other ensembles and control parameters can be treated on the same footing. For example, the PA simulations of hard-sphere mixtures reported in Ref.~\cite{callaham:17} use packing fraction as the control parameter.

    \subsection{General PA framework}\label{sec:ALG:alg}
    
        In order to capture the full potential of PA and weighted averages, the algorithm described below applies to (almost) arbitrary target distributions $\rho_\beta$ and hence generalizes the notation with respect to the PA literature such as Refs.~\cite{machta:10a, wang:15a, weigel:21}.
        At the end, we explicitly discuss the case of the canonical ensemble that is also realized in the numerical simulations discussed in Sec.~\ref{sec:num}.
        
        Let $\beta_0, \ldots, \beta_f$ be an annealing schedule and suppose that the respective target distributions are known up to constants,
        \begin{equation}
             \rho_{\beta_i}(\gamma) = \frac{v_i(\gamma)}{C_i}\qquad \forall \gamma \in \Gamma.
        \end{equation}
        The sequence $\rho_{\beta_0}$, $\rho_{\beta_1}$, $\ldots$ must be chosen such that the overlaps $\int \rho_{\beta_i}\rho_{\beta_{i-1}}\, \text{d} \gamma$ are sufficiently large~\cite{barash:16,weigel:21}.
        After selecting a \emph{target population size} $R \gg 1$ one may proceed as follows:
        
        \begin{enumerate}[(i)]
            \item Draw $R_0 \coloneqq R$ independent configurations from $\rho_{\beta_0}$. Put $i=1$. 
            
            \item \label{alg:step:ii} For all $1 \leq j \leq R_{i-1}$, calculate the \emph{scaled weight ratio} in the state $\gamma_{i-1}^{(j)}$ sampled by replica $j$ at $\beta_{i-1}$,
            % use alternative symbol for tau here, to avoid confusion with \hat estimators appearing later
            \begin{equation}\label{eq:tau}
                \tau_i^{(j)} \coloneqq \frac{R}{R_{i-1}} \frac{1}{Q_i} \frac{v_i(\gamma_{i-1}^{(j)})}{v_{i-1}(\gamma_{i-1}^{(j)})},
            \end{equation}
            where the following normalization is used,
            \begin{equation}\label{eq:Q}
                Q_i \coloneqq \frac{1}{R_{i-1}} \sum_{j=1}^{R_{i-1}} \frac{v_i(\gamma_{i-1}^{(j)})}{v_{i-1}(\gamma_{i-1}^{(j)})}.
            \end{equation} 
            
            \item \label{alg:step:iii}
            Resampling: The number $r_i^{(j)}$ of descendants of replica $j$ from $\beta_{i-1}$ to $\beta_i$ is a non-negative integer drawn from a distribution with mean $\tau_i^{(j)}$.
            For instance, one may use nearest-integer resampling~\cite{wang:15a},
            \begin{equation}
                r_i^{(j)} = 
                \begin{cases}
                    \ceil{ \tau_i^{(j)}} & \text{with probability}~\tau_i^{(j)} -  \floor{ \tau_i^{(j)}},\\
                    \floor{ \tau_i^{(j)}} & \text{otherwise}.
                \end{cases}
            \end{equation}
            Herein, $\ceil{x}$ ($\floor{x}$) refers to the smallest (largest) integer greater (less) than or equal to $x$, respectively.
            Update the population size $R_{i} \coloneqq \sum_{j} r_i ^{(j)}$. 

            \item \label{alg:step:iv} Apply an equilibration routine causing the resampled population to approach $\rho_{\beta_i}$.
            In order for Eq.~\eqref{eq:tau} to be well-defined at $\beta_{i+1}$, ensure that no forbidden configurations are sampled, i.e.
            \begin{equation}
                \rho_{\beta_i}(\gamma_{i}^{(j)})>0
            \end{equation}
            for every replica $j$ in the state $\gamma_{i}^{(j)}$ at $\beta_i$.
            
            \item \label{alg:step:v} Measure observables (details are given in Sec.~\ref{sec:MO}).
            
            \item If $\beta_i < \beta_f$, increment $i$ and go to step \ref{alg:step:ii}.
        \end{enumerate}
        Note that the expected number of copies created of replica $j$ during step \ref{alg:step:iii} only differs from $\rho_{\beta_{i}}/\rho_{\beta_{i-1}}$ evaluated in the state of replica $j$ by a replica-independent factor.
        The term $R/(R_{i-1}Q_i)$ in Eq.~\eqref{eq:tau} assures that the average population size $R_i$ at $\beta_i$ equals $R$, although small fluctuations occur based on the variance of the (nearest-integer) resampling scheme in step \ref{alg:step:iii}~\cite{gessert:22a}.
        
        In case of the canonical ensemble, we employ annealing schedules of the form $0=\beta_0< \beta_1< \ldots < \beta_f$ and the weight function becomes $v_i(\gamma)=\exp[-\beta_i H(\gamma)]$, where $H$ is the system's Hamiltonian.
        Consequently, one has
        \begin{align}
            \tau_i^{(j)} \coloneqq \frac{R}{R_{i-1}} \frac{1}{Q_i} &\exp \left[ - (\beta_i - \beta_{i-1})H(\gamma_{i-1}^{(j)})\right],\label{eq:tauBoltz}\\
            Q_i \coloneqq \frac{1}{R_{i-1}} \sum_{j=1}^{R_{i-1}} &\exp \left[ - (\beta_i - \beta_{i-1})H(\gamma_{i-1}^{(j)})\right],\label{eq:QBoltz}
        \end{align}
        recovering established algorithms such as Refs.~\cite{wang:15a,weigel:21}.

    \subsection{Sources of bias and asymptotics}\label{sec:ALG:bias}
    
        Following the algorithm above, the initial population is an unbiased sample from $\rho_{\beta_0}$ and only statistical fluctuations are present at $\beta_0$.
        For subsequent annealing steps however, finite population sizes and finite time spent in equilibration routines also cause systematic errors~\cite{weigel:21}.
        
        In fact, the resampling step \ref{alg:step:iii} reduces the number of independent replicas and introduces correlations.
        Moreover, fluctuations due to nearest-integer resampling are only \emph{conditionally} unbiased, i.e., resampling is only accurate ``on average'' given that the population at $\beta_{i-1}$ is perfectly equilibrated.
        
        In the limit $R \to \infty$, step \ref{alg:step:iii} almost surely transforms $\rho_{\beta_{i-1}}$ into $\rho_{\beta_i}$ and no equilibration routines are needed as both systematic and statistical errors vanish~\cite{wang:15a}.
        Unsurprisingly, if equilibration routines such as MCMC algorithms receive an infinite amount of resources, populations represent $\rho_{\beta_i}$ regardless of the resampling behavior.
        In practice, correlation-inducing resampling and decorrelating MCMC routines perform well in conjunction, since they naturally alleviate each other's shortcomings.

\section{Models and observables}\label{sec:MO}

  PA is a fairly general approach and our analytical results as well as the main conclusions from the numerical simulations are model independent.
  However, the practical assessment of the effect of weighted averaging relies on simulations of concrete models. To cover a wide range of practically relevant behaviors, we test our predictions and analyze in detail systematic and statistical errors for weighted averages of simulations for the Ising ferrogmagnet (FM) and the  Edwards-Anderson Ising spin glass (SG), both in two dimensions. The former is an example of a simple model with a continuous phase transition while the latter is a problem with metastability and a complex free-energy landscape. 
  
    \subsection{Ising ferromagnet and spin glass}\label{sec:MO:Models}

        Both systems are studied on square lattices of linear size $L$ and each state $\gamma = (s_1,\ldots,s_N)$ of these models corresponds to a choice of $N=L^2$ Ising spins $s_n = \pm 1$.
        Thus, $\Gamma$ has cardinality $2^N$.
        Only nearest neighbors $m \neq n$, denoted $\langle m,n \rangle$, are allowed to interact directly via coupling constants $J$ resp.\ $J_{mn}$, and periodic boundary conditions are employed.
        In the absence of external fields, the Ising FM is described by the Hamiltonian
        \begin{equation}
                 H_{\mathrm{FM}}(\gamma) \coloneqq - J\sum_{\langle m , n \rangle} s_m s_n,\label{eq:H_ferro}
        \end{equation}
        where in the following we set $J=1$. This form is generalized for the SG model to read
        \begin{equation}
                 H_{\mathrm{EA}}(\gamma) \coloneqq - \sum_{\langle m , n \rangle} J_{mn} s_m s_n.\label{eq:H_SG}
        \end{equation}
        Here, $J_{mn}$ are quenched random variables drawn from $\{\pm 1\}$ uniformly and independently. In general one is interested in the disorder average of observables over such coupling realizations $\mathcal{J} \coloneqq \{J_{mn}: \langle m,n \rangle\}$. For the purposes of our study, however, it is also meaningful to consider individual realizations such as the hardest instance encountered. Without further qualification, in the following $H$ or $H(\gamma)$ stands for either of the two Hamiltonian functions, Eqs.~\eqref{eq:H_ferro} or \eqref{eq:H_SG}.
        Recall that the critical inverse-temperature of the above two-dimensional Ising FM in the thermodynamic limit is given by~\cite{onsager:1944, kaufman:1949}
        \begin{equation}
            \beta_c=\frac{1}{2J}\ln(1+\sqrt{2}),
        \end{equation}
        in particular $\beta_c\approx 0.4407$ in our case where $J=1$.
        
    \subsection{Ensemble averages}\label{sec:MO:EnsmblAvrg}
    
        Some of the most fundamental quantities in statistical physics are ensemble averages of observables $\mathcal{O}$ at $\beta_i$,
        \begin{equation}\label{eq:configurational}
            \langle \mathcal{O} \rangle_{\beta_i} \coloneqq \int_{\Gamma} \mathcal{O}(\beta_i, \gamma)\rho_{\beta_i}(\gamma)\, \text{d}\gamma.
        \end{equation}
        Approximation schemes such as Monte Carlo simulations strive to estimate such expectation values. PA populations are close to samples drawn from the respective equilibrium distribution $\rho_{\beta_i}$, so a natural estimator for Eq.~\eqref{eq:configurational}
        during step \ref{alg:step:v} is the population average~\cite{machta:10a}
        \begin{equation}\label{eq:PopAvrg}
            \widehat{\mathcal{O}}_i \coloneqq \frac{1}{R_i} \sum_{j=1}^{R_i} \mathcal{O}(\beta_i, \gamma_{i} ^{(j)}),
        \end{equation}
        where $\gamma_i^{(j)}\in \Gamma$ refers to the state of replica $j$ at $\beta_i$.
        Let $\widehat{\rho}_i(\gamma)$ be the empirical density obtained from a single population at $\beta_i$
        , i.e.~\footnote{Note the slight abuse of notation here as addition on $\Gamma$ is not necessarily well defined. What we only mean to state is that $\delta(\gamma-\gamma_i^{(j)})$ is an object such that for a continuous function $f$ on $\Gamma$ one has \[\int_\Gamma f(\gamma)\,\delta(\gamma-\gamma_i^{(j)}) \mathrm{d}\gamma = f(\gamma_i^{(j)}).\]},
        \begin{equation}
          \widehat{\rho}_i(\gamma) = \frac{1}{R_i}\sum_{j=1}^{R_i} \delta(\gamma-\gamma_i^{(j)}).
        \end{equation}
        Then Eq.~\eqref{eq:PopAvrg} is equivalent to
        \begin{equation}
            \widehat{\mathcal{O}}_i = \int_{\Gamma}\mathcal{O}(\beta_i, \gamma) \widehat{\rho}_i(\gamma) \, \text{d}\gamma.
        \end{equation}
        Thus, we use the following estimators for the (internal) energy and magnetization per spin
        \begin{align}
            \widehat{e}_i &\coloneqq \frac{1}{NR_i} \sum_{j=1}^{R_i} H(\gamma_i ^{(j)}),\label{eq:e}\\
            \widehat{m}_i &\coloneqq \frac{1}{N R_i} \sum_{j=1}^{R_i} \left| \sum_{n=1}^N s_n(\gamma_i^{(j)}) \right|.\label{eq:m}
        \end{align}
        
        More generally, if observables can be expressed in terms of functions of ensemble averages, one may derive the appropriate PA estimator by substituting population averages instead.
        In this manner, heat capacity and susceptibility per spin can be measured in step \ref{alg:step:v} using
        \begin{align}
            \widehat{c}_i &\coloneqq \beta_i^2 N \left(\widehat{e_i ^2} - (\widehat{e}_i) ^2\right)\label{eq:c},\\
            \widehat{\chi}_i &\coloneqq \beta_i N \left(\widehat{m_i ^2} - (\widehat{m}_i) ^2\right)\label{eq:chi},
        \end{align}
        where $ \widehat{e_i ^2}$ and $\widehat{m_i ^2}$ are defined similarly to Eq.~\eqref{eq:e} and~\eqref{eq:m} with squared summands.
        Quantities involving $m$ are only computed in the FM case.
        With regards to the Ising SG, all definitions given here rely on an implicit choice of disorder realization $\mathcal{J}$.
        Spin overlap measurements are discussed separately in Sec.~\ref{sec:MO:Overlap}. 
        
    \subsection{Free energy}\label{sec:MO:F}
    
        PA naturally allows for measurements of the potential associated to the considered ensemble. This is most easily seen for the free energy $F(\beta)$ in the canonical ensemble (but see Refs.~\cite{callaham:17,rose:19} for the microcanonical case).
        Based on the \emph{partition function}
        \begin{equation}\label{eq:Z_def}
            Z(\beta) \coloneqq \int_\Gamma \exp[-\beta H(\gamma)] \, \text{d}\gamma,
        \end{equation}
        it holds
        \begin{equation}\label{eq:F_def}
            F(\beta) \coloneqq -\frac{1}{\beta} \ln Z(\beta).
        \end{equation}
        Thus, $F$ admits the following telescopic expansion~\cite{machta:10a,wang:15a}
        \begin{equation}\label{eq:F_tele}
            - \beta_i F(\beta_i) = \sum_{k=1}^i \ln \frac{Z(\beta_k)}{Z(\beta_{k-1})} + \ln Z(\beta_0).
        \end{equation}
        The ratio of partition functions $Z(\beta_i)/Z(\beta_{i-1})$ is exactly the state-independent constant relating $\rho_{\beta_i}/\rho_{\beta_{i-1}}$ to $v_i/v_{i-1}$ and it is naturally estimated by Eq.~\eqref{eq:QBoltz}~\cite{machta:10a},
        \begin{align}
            \frac{Z(\beta_i)}{Z(\beta_{i-1})}& 
            = \frac{1}{Z(\beta_{i-1})} \int_\Gamma \exp\left[-\beta_i H(\gamma)\right] \, \text{d}\gamma \nonumber \\ 
            &= \int_\Gamma \exp\left[-(\beta_i - \beta_{i-1})H(\gamma)\right] \rho_{\beta_{i-1}}(\gamma) \, \text{d}\gamma \nonumber \\
            &= \langle\exp\left[-(\beta_i - \beta_{i-1})H(\gamma)\right] \rangle_{\beta_{i-1}}\approx Q_i.\label{eq:Zratio}
        \end{align}
        More generally, a similar calculation invoking Eq.~\eqref{eq:Q} shows $C_i/C_{i-1} = \langle v_i/ v_{i-1} \rangle_{\beta_{i-1}} \approx Q_i$, given that $\mathrm{supp}(\rho_{\beta_i}) \subseteq \mathrm{supp}(\rho_{\beta_{i-1}})$, i.e. given that the ratio $v_i/ v_{i-1}$ is defined.
        Together, Eqs.~\eqref{eq:F_tele} and~\eqref{eq:Zratio} yield the free-energy estimator at $\beta_i$~\cite{machta:10a},
        \begin{equation}\label{eq:F}
            - \beta_i \widehat{F}_i \coloneqq \sum_{k=1}^i \ln Q_k + \ln Z(\beta_0),
        \end{equation}
        which can be obtained without further computational expense from step \ref{alg:step:ii}.
        In the Ising cases above, one has $Z(\beta_0)=Z(0)=N \ln 2$.
        The division of $\widehat{F}_i$  by $N$ leads to the free-energy per spin estimator, denoted by $\widehat{f}_i$.
        If $ \ln Z(\beta_0)$ is unknown, e.g., if $\beta_0 > 0$, only free-energy differences can be obtained.
        
        Note that $\widehat{F}$ directly incorporates information of the whole annealing process, in contrast to the single-temperature estimators in Eq.~\eqref{eq:e} to \eqref{eq:chi}.
        While this leads to smooth estimates, any bias ``picked up'' throughout the annealing process is still present at subsequent temperatures.
        For the $d$-dimensional Ising FM with constant coordination number $z$ one has $f \to -z/2$ in the limit $\beta \to \infty$.
        This is sufficient to derive that bias decays proportional to $\beta^{-1}$ for $\beta \to \infty$, as we show in App.~\ref{app:F_bias}.

    \subsection{Spin overlap}\label{sec:MO:Overlap}
    
        For the spin glass the magnetization of Eq.~\eqref{eq:m} does not provide an order parameter and, instead, we consider the 
        spin overlap of two replicas with the same coupling configuration (Parisi overlap parameter) \cite{mezard:book},
        \begin{equation}
            q_\mathcal{J}(\gamma, \gamma') \coloneqq \frac{1}{N}\sum_{n=1}^{N} s_n(\gamma)s_n(\gamma').
        \end{equation}
        The quantity of main interest then is the probability of finding a specific overlap $q$, i.e.,
        \begin{equation}
            \label{eq:overlap-distribution}
            P_\mathcal{J}(q) = \int_{\Gamma} \int_{\Gamma} \delta(q_\mathcal{J}(\gamma,\gamma')-q)\, \rho_{\beta}(\gamma)\rho_{\beta}(\gamma')\, \text{d} \gamma \text{d} \gamma'.
        \end{equation}
        Here, we make the dependence on the disorder realization $\mathcal{J}$ explicit in order to clearly distinguish it from the disorder average
        \begin{equation}\label{eq:Pq_disAvrg}
            P(q) = \left[P_\mathcal{J}(q)\right]_\mathrm{av}.
        \end{equation}
        A scalar order parameter can be constructed by considering the mean absolute value of $q$,
        \begin{equation}\label{eq:ensembleAvrg_absQ}
            \langle |q| \rangle_{\beta,\mathcal{J}} = \int P_\mathcal{J}(q) |q| \, \text{d} q,
            \hspace*{0.3cm}
            \langle |q| \rangle_{\beta} = \int P(q) |q| \, \text{d} q.
        \end{equation}
        Measuring the distribution \eqref{eq:overlap-distribution} of $q$ in a simulation requires independent pairs $(\gamma, \gamma')$, thus usually doubling the required computational effort.
        For instance, it is common to use two separate parallel tempering runs to obtain configurations completely independent from each other~\cite{wang:15a, kumar:20}. 
        
        In the following paragraphs, we discuss different approaches to measure the spin-overlap distribution $P_\mathcal{J}(q)$ in PA without having to run multiple simulations.
        Some of these approaches are based on the concept of  \emph{families}, which are the descendants of a single replica in the initial population~\cite{wang:15a}.
        Replicas from different families evolve independently throughout the annealing process, except for the resampling step \ref{alg:step:iii} where normalizing by $Q_i$ allows replicas to influence each other's progeny.
        Thus, family sizes may be correlated, although replicas from different families are independent.
        For ease of implementation, we switch to zero-based indexing for the remainder of this section.

        \subsubsection{Independent pairings from permutations}
        
            Wang \emph{et~al.}~\cite{wang:15a} proposed a method to obtain $R_i$ spin overlap values at $\beta_i$ in worst-case time complexity $\mathcal{O}(R_i ^2)$ using every replica exactly twice.
            To ensure independence, only pairs from different families are considered, i.e., a permutation
            \begin{equation}
                \pi^* = (\pi^*(0), \ldots, \pi^*(R_i-1))\eqqcolon(\pi^*_0, \ldots, \pi^*_{R_i-1})
            \end{equation}
            is needed, satisfying that $k$ and $\pi^*_k$ belong to different families for all $0 \leq k \leq R_i-1$.
            As long as family sizes are below $R_i/2$, such $\pi^*$ exists and can be found by drawing a random initial permutation $\pi$ and repeating the following:
            Let $k$ be the smallest index with an ``incestuous'' pairing $\pi_k$ and use the short-hand notation $k+l$ for $(k+l)~\text{mod}~R_i$.
            Iterate through $\pi_{k+1}, \pi_{k+2}, \ldots$ until an element $\pi_{k+l}$ is found such that $\pi_{k+l}$ is not in the family of $k$ and $\pi_k$ is not in the family of $k+l$.
            Then swap $\pi_k$ and $\pi_{k+l}$ to lower the number of incestuous pairs.
            Since family sizes do not exceed $R_i/2$, a suitable transposition is found after at most $R_i-1$ attempts which ensures termination.
            Although the algorithm has quadratic worst-case time complexity in $R_i$, it was claimed to be close to linear in practice~\cite{wang:15a}.
            
            We tested this approach and could not find significant deviations from running two independent simulations with regards to bias.
            This can be seen, for example, in the upper panel of Fig.~\ref{fig:qHist_WangComparison} showing data for the ``hardest'' instance encountered for the two-dimensional Ising SG in a sense to be described in Sec.~\ref{sec:Meth:rhot}.
            The figure is discussed in greater detail below.
            
            Note that the algorithm described here does not transform a uniform distribution of initial permutations into a uniform distribution on the set of ``non-incestious'' $\pi^*$, which can be easily checked numerically. 
            We expect this not to be problematic for the PA use case as long as there is no prescribed order within families, e.g., energetically ascending.
            Numerically, we see that the introduction of a random search pattern can restore this uniformness property if needed.
            That is, another random permutation $\sigma$ may be drawn at the start,
            mismatches may be checked along the positions $\sigma_1, \sigma_2, \ldots$ and the sequence $(\pi \circ \sigma)_{i+1}, (\pi \circ \sigma)_{i+2}, \ldots$  used to find a transposition for a mismatch at $k=\sigma_i$.
            This search process is stopped if $(\pi \circ \sigma)_{i+l}$ is found such that $(\pi \circ \sigma)_{i+l}$ is not in the family of $\sigma_i$ and $(\pi \circ \sigma)_i$ is not in the family of $\sigma_{i+l}$.
            The original algorithm is recovered by fixing $\sigma=\text{id}$.
            
            We struggled to parallelize the approach by Wang \emph{et al.}, however, resulting in a serial bottleneck in the optimized GPU code of Ref.~\cite{barash:16} that our numerical simulations are based on.
            There are also implementation-independent drawbacks.
            Due to the strong sample-to-sample fluctuations that are typical for spin-glass systems (see, e.g., Ref.~\cite{wang:15a}), the existence of families larger than $R_i/2$ can often not be ruled out beforehand and if such populations are encountered it is unclear how to proceed.
            Possible choices include employing the present incestuous permutation, omitting the $q$ measurement or terminating the simulation entirely.
            Depending on the frequency of spin overlap measurements throughout the annealing process, it may be likely to encounter the same problem again at subsequent annealing steps in the first two cases.
            The last option poses the risk of rejecting simulations sampling rare low-energetic states, thereby introducing a new source of bias.

        \subsubsection{Independent pairings from index shifts}
        
            \begin{figure}
                \includegraphics[width=\linewidth]{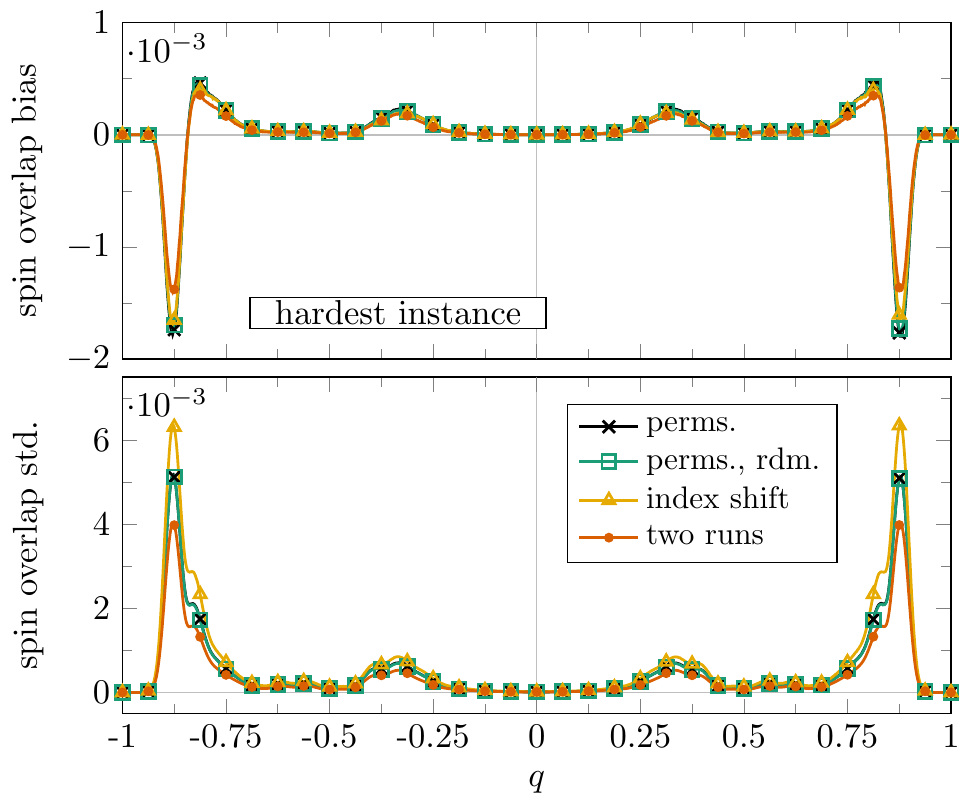}
                \caption{\label{fig:qHist_WangComparison}
                Systematic errors (top) and standard deviations (bottom) in the spin overlap histogram of the hardest $L=32$ Ising SG instance encountered at $\beta=2.4$ employing $\theta=10$ Metropolis sweeps to a target population size $R=2\times 10^4$.
                Compared are the permutation method by Wang \emph{et al.}~\cite{wang:15a}, the modified version of the permutation method with a random search pattern, as well as our suggestion of using index shifts and forming pairs from two independent simulations.
                In the two latter cases we also excluded measurements from populations where a family exceeded $R_i/2$ for a clearer comparison.
                The number of repeated simulations entering this analysis is shown in Table~\ref{tab:qHist_Term}.
                For easier readability, only every 32nd histogram value carries a symbol.
                }
            \end{figure}
            
            \begin{table}[b]
                \centering
                \caption{\label{tab:qHist_Term}
                Number of simulations entering the comparison of $q$ measurements in Fig.~\ref{fig:qHist_WangComparison}.
                Runs where one family at $\beta=2.4$ exceeded the size of $R_i/2$ were excluded and counted towards the fraction in the last column.
                One repetition of ``two runs'' corresponds to two independently simulated populations in one GPU program to parallelize the calculation of $q$.
                Since such a repetition is only used if none of the families in two populations exceeds $R_i/2$, the excluded fraction in the last row is larger.
                }
                \begin{ruledtabular}
                \begin{tabular}{cccccc}
                    method & repetitions & included & excluded fraction\\ % included in percent
                    \colrule
                    Wang        & $5 \times 10^4$ & $42308$  & 15.4\%   \\
                    Wang, rdm   & $5 \times 10^4$ & $42308$  & 15.4\%   \\
                    Index shift & $5 \times 10^4$ & $42092$  & 15.8\%   \\
                    Two runs    & $10^5$          & $71374$  & 28.6\%   \\
                \end{tabular}
                \end{ruledtabular}
            \end{table}

            As a simple alternative, we considered computing $\floor{R_i/2}$ spin overlaps from a population of size $R_i$ by choosing pairs with distance $\floor{R_i/2}$ in replica index space.
            This is based on the fact that our implementation deliberately places resampled copies next to each other, resulting in indices of family members being contiguous.
            Hence, pairing replica $j$ with $j + \floor{R_i/2}$ for $0 \leq j < \floor{R_i/2}$ avoids the problem of dependence under the same assumption that family sizes are below $R_i/2$.
            This algorithm is clearly simpler and faster than the permutation method, but the potential price of this speedup are ``blocks'' of similar overlap values whenever the employed equilibration routine is insufficient.
            However, this should only increase statistical errors, since there is no prescribed relation between a certain family and the families placed at a distance of $\floor{R_i/2}$.
            
            We see this for example in Fig.~\ref{fig:qHist_WangComparison} showing the ``hardest'' instance encountered at $\beta=2.4$, as explained in Sec.~\ref{sec:Meth:rhot}.
            There is virtually no difference in systematic errors between using index shifts and the Wang \emph{et al.} permutation approach while a  minor increase in statistical errors is present.
            Introducing a random search pattern for transpositions does not change bias or statistical fluctuations in our implementation.
            If an ordering of replicas is imposed, however, this approach would probably yield better results.
            Of course, the best estimates are obtained by forming pairs from independent runs at the price of doubling the required computational work.
            To get a cleaner comparison to the approach of Wang {\em et al.\/} in Fig.~\ref{fig:qHist_WangComparison}, we also rejected measurements if a family was larger than $R_i/2$ while using index shifts or two independent runs.
            The number of included and excluded simulations is specified in Table~\ref{tab:qHist_Term}.
            Note that $\theta=10$ was chosen to provoke deviations in the comparison through insufficient equilibration, which results in a relatively large fraction of simulations exceeding the family size constraint.
            
            Another advantage of using index shifts is that it preserves the ``locality'' of correlations, thereby enabling the blocking analysis introduced in Ref.~\cite{weigel:21}.
            That is, if correlations are localized in replica index space, correlated $q$ measurements using index shifts will still posses this property and an effective population size $R_{\mathrm{eff}}(q)$ based on the quality of spin overlap measurements as well as statistical errors of estimators can be computed from within a single PA run.
            
            In conclusion, we form replica pairs from index shifts as it provides the best results for GPU runtime in our case.
            More details on the implementation of $q$ measurements and reference solutions are given in Sec.~\ref{sec:Metho:Sol}.

\section{Weighted averages}\label{sec:WA}

    The focus of the present work is on the analysis of ways to reduce both systematic and statistical errors in PA through data from $M$ independent simulations.
    Machta~\cite{machta:10a} first recognized this possibility and coined the term \emph{weighted average}, motivated by the appropriate formula for particularly ``simple'' observables.
    He claimed that bias vanishes in the limit $M \to \infty$.
    These ideas have been employed in several subsequent publications~\cite{wang:15a, amey:18, callaham:17, rose:19}.
    The justification for this approach, however, remained to be largely based on analogies related to a theoretical version of PA called \emph{unnormalized population annealing} (uPA)~\cite{machta:10a, wang:15a} (but see Ref.~\cite{weigel:21} for an alternative line of argument).
    
    To give a self-contained presentation of the established theory behind weighted averaging, we start by introducing the general arguments leading to the weight functions in Sec.~\ref{sec:WA:ideas}.
    The resulting weighted averaging formulas are presented in Sec.~\ref{sec:WA:conf} and Sec.~\ref{sec:WA:F}, before we turn to a rigorous result on the convergence of weighted averages in the absence of equilibration routines in Sec.~\ref{sec:WA:proof}.
    Hereafter, appropriate estimators for observables defined in terms of central moments such as the heat capacity and susceptibility are derived.
    Lastly, weighted averages for the spin-overlap distribution and the variance of free-energy weights are discussed in Sec.~\ref{sec:WA:q} and~\ref{sec:WA:w}.
    
    \subsection{Key ideas and free-energy weights}\label{sec:WA:ideas}
    
        Weighted averaging exploits the existence of populations, which can be ``merged'' to gain a larger sample.
        However, this cannot be done trivially, since simply adding populations from independent runs corresponds to adding identically distributed quantities and therefore preserves systematic errors.
        To derive the correct way of merging populations, one can use the following reasoning~\cite{machta:10a, wang:15a, weigel:21}:
        
        Consider a slight modification to the algorithm described in Sec.~\ref{sec:ALG:alg} applied to the canonical ensemble, where the desired number of copies of replica $j$ in Eq.~\eqref{eq:tauBoltz} is solely defined as the exponential expression, i.e., without the prefactor $R/R_{i-1}Q_i$.
        Thereby, the primary tool of population size control is removed and, depending on the energy reference point, resampling can drastically increase or decrease the number of replicas, rendering the uPA scheme impractical~\cite{wang:15a}.
        At the same time, the normalization presents the only interaction between competing families.
        As a consequence of its removal, it is impossible to tell whether a single uPA run was initialized with states $\gamma_1, \ldots, \gamma_R$ at $\beta_0$ and therefore produced a collection of surviving families at $\beta_i$ or if $R$ uPA runs indexed $1 \leq r \leq R$ were initialized in single states $\gamma_r$ and the surviving replicas in every run at $\beta_i$ unified trivially.
        As this argument generalizes to any partition of the initial population in uPA, we obtain a convenient property.
        If we measure the population average $\widehat{\mathcal{O}}_i$ from Eq.~\eqref{eq:PopAvrg} in $M$ independent uPA runs, the resulting estimates $\widehat{\mathcal{O}}_i^{(1)}, \ldots, \widehat{\mathcal{O}}_i^{(M)}$ should be combined via
        \begin{equation}\label{eq:weighted_uPA}
            \sum_{m=1}^M \frac{R_i^{(m)}}{R_i^{(1)} + \ldots +R_i^{(M)}} \widehat{\mathcal{O}}_i^{(m)} \eqqcolon \sum_{m=1}^M \tilde{w}_i^{(m)} \widehat{\mathcal{O}}_i^{(m)}.
        \end{equation}
        If the independent uPA runs are initialized with population sizes $R^{(1)}, \ldots, R^{(M)}$, this estimator is \emph{equivalent} to measuring $\widehat{\mathcal{O}}_i$ in a uPA simulation with initial population size $R=R^{(1)}+ \ldots +R^{(M)}$.
        Thus, it is unbiased in the limit $M \to \infty$~\cite{wang:15a} since this corresponds to $R \to \infty$.
        
        In view of Eq.~\eqref{eq:weighted_uPA}, the key idea is to estimate the population size which a standard PA run would have reached in the unnormalized setting and use this number as a weight for the population~\cite{machta:10a, wang:15a}.
        Since the expected population size after unnormalized resampling at $\beta_{k-1}$ is $R_{k-1}Q_k$ instead of $R$, multiplying the ratios $R_{k-1}Q_k/R$ for all $k \leq i$ yields an estimate for the ratio of uPA and standard PA population sizes at $\beta_i$.
        Thus, independent PA runs with identical annealing schedules, equilibration algorithms, and target population sizes $R$ should be weighted against each other at $\beta_i$ according to
        \begin{equation}\label{eq:weight_derivation}
            \prod_{k=1}^{i} \frac{R_{k-1}}{R} Q_k = \frac{1}{Z_0} \prod_{k=1}^{i} \frac{R_{k-1}}{R} \exp(-\beta_i \widehat{F}_i),
        \end{equation}
        where we have used Eq.~\eqref{eq:F}. Consequently, as is shown based on somewhat different arguments in Ref.~\cite{weigel:21}, data from runs $1 \leq m \leq M$ should carry the following \emph{free-energy weight},
        \begin{equation}\label{eq:appropriateWeight}
            w_i^{(m)} \coloneqq \frac{ R_i^{(m)} \prod_{k=1}^i (R_{k-1}^{(m)}/R^{(m)}) \exp(-\beta_i \widehat{F}_i ^{(m)} ) }{ \sum_{m'} R_i^{(m')} \prod_{k=1}^i (R_{k-1}^{(m')}/R^{(m')}) \exp(-\beta_i \widehat{F}_i ^{(m')} ) }.
        \end{equation}
        Additionally, the \emph{simplified free-energy weight} is considered, which we expect to yield similar results~\cite{machta:10a, wang:15a, weigel:21},
        \begin{equation}\label{eq:simpleWeight}
            \underline{w}_i^{(m)} \coloneqq \frac{ R_i^{(m)} \exp(- \beta \widehat{F}_i^{(m)}) }{ \sum_{m=1}^M R_i^{(m)} \exp(- \beta \widehat{F}_i^{(m)}) }.
        \end{equation}
        This form is exact for the case of constant population sizes during the anneal~\cite{weigel:21}, but it also provides a reasonable approximation for not too large relative fluctuations in population size.
        
        A potential flaw in this argumentation is that the estimation quality of the hypothetical unconstrained population size of a de facto constrained population remains rather unclear.
        After all, the method appears to be based on treating PA observables as uPA observables whilst they are differently distributed and hence behave differently during resampling.

    \subsection{Configurational estimators}\label{sec:WA:conf}
    
        The most natural use of the free-energy weights is for computing weighted averages for ``elementary'' observables $\mathcal{O}$ of the simulation, for example the energy or magnetization. In this case, the appropriate weighted estimator $\W{\widehat{\mathcal{O}}_i}$ for the population average $\widehat{\mathcal{O}}_i$ from Eq.~\eqref{eq:PopAvrg} is the \emph{configurational weighted average}
        \begin{equation}\label{eq:W_conf}
            \W{\widehat{\mathcal{O}}_i} \coloneqq \sum_{m=1}^M w_i^{(m)} \widehat{\mathcal{O}}_i^{(m)}.
        \end{equation}
        Machta~\cite{machta:10a} claimed that $\W{\widehat{\mathcal{O}}_i}$ is asymptotically unbiased with respect to $M \to \infty$, in view of the arguments given above.
        The similarly defined estimator with $\underline{w}$ substituted for $w$ is denoted as $\uW{\widehat{\mathcal{O}}_i}$.
        
        For more general observables such as, for instance, the free energy, specific heat and susceptibility, this basic weighting scheme is \emph{not} suitable \cite{machta:10a,callaham:17} and appropriate modifications are given in Secs.~\ref{sec:WA:F} and \ref{sec:WA:moments} below. As it stands, Eq.~\eqref{eq:W_conf} only applies to \emph{configurational} estimators, where we call an estimator $\widehat{\mathcal{O}}_i$ configurational, if it is defined in terms of Eq.~\eqref{eq:PopAvrg}, where $\mathcal{O}(\beta_i, \gamma)$ can be calculated \emph{without} information on the distribution $\rho_{\beta_i}$.
        More generally, we refer to asymptotically unbiased estimators with respect to $M \to \infty$ as \emph{weighted estimators} and call the weighted estimator for configurational quantities \emph{configurational weighted average}.
        
        We also note the following useful property: Suppose that configurational weighted averages are asymptotically unbiased $\W{\widehat{\mathcal{O}}_i}\to \langle \mathcal{O} \rangle_{\beta_i}$ and consider a function $g$ that is continuous around $\langle \mathcal{O} \rangle_{\beta_i}$, then it holds that $g(\W{\widehat{\mathcal{O}}_i}) \to g(\langle \mathcal{O} \rangle_{\beta_i})$. Hence, we immediately know how to deal with continuous functions of configurational estimators.
    
    \subsection{Free energy}\label{sec:WA:F}
    
        Thus, in view of Eq.~\eqref{eq:F}, a reasonable \emph{weighted free-energy estimator} is~\cite{machta:10a}
        \begin{equation}\label{eq:W_F}
            - \beta_i \W{\widehat{F}_i} \coloneqq \sum_{k=1}^i \ln \W{Q_k} + \ln Z(\beta_0),
        \end{equation}
        where we have to take into account that the configurational estimator $Q_k$ is evaluated at $\beta_{k-1}$, i.e.,
        \begin{equation}
            \W{Q_k} = \sum_{m=1}^M w_{k-1}^{(m)} Q_k^{(m)}.
        \end{equation}
        The weighted estimator \eqref{eq:W_F} takes a particularly simple form in case of constant population size during the annealing process (for example using multinomial resampling).
        If one also uses identical initial population sizes in the runs to be combined, the weights of Eq.~\eqref{eq:appropriateWeight} simplify to
        \begin{equation}
            w_i^{(m)} = \frac{\exp(-\beta_i \widehat{F}_i ^{(m)} )}{\sum_{m'=1}^M \exp(-\beta_i \widehat{F}_i ^{(m')} )} = \underline{w}_i ^{(m)}.
        \end{equation}
        Substituting these weights into Eq.~\eqref{eq:W_F} leads to a telescopic expression which resolves to~\cite{machta:10a}
        \begin{equation}\label{eq:W_F_machta}
            - \beta_i \W{\widehat{F}_i} = \ln \left[ \frac{1}{M} \sum_{m=1}^M \exp(-\beta_i \widehat{F}_i ^{(m)} ) \right].
        \end{equation}
        That is, \textit{weighted averaging is performed on the level of partition functions}~\cite{machta:10a}.
        In the more general case, Eq.~\eqref{eq:appropriateWeight} and \eqref{eq:simpleWeight} do not obey this telescopic property, which results in rather lengthy explicit expressions for $\W{\widehat{F}_i}$ and $\uW{\widehat{F}_i}$ in terms of $\widehat{F}$.
        Still, they remain to be incremental with respect to subsequent annealing steps, allowing them to be computationally cheap.

    \subsection{Convergence to equilibrium distribution}\label{sec:WA:proof}
        
        Suppose that $\Gamma$ is finite and a PA algorithm analogous to Sec.~\ref{sec:ALG:alg} is applied satisfying the following conditions.
        \begin{enumerate}[(a)]
        
            \item \label{item:proof:a} $\mathbb{E}[\widehat{\rho_0}(\gamma)] = \rho_{\beta_0}(\gamma)\quad \forall \gamma \in \Gamma$.
            
            \item \label{item:proof:b} Regions in $\Gamma$ to which the target distributions attribute positive probability are not expanding throughout the anneal, i.e., for all $i$ it holds
            \begin{equation}
                \text{supp}(\rho_{\beta_i}) \subseteq \text{supp}(\rho_{\beta_{i-1}}).
            \end{equation}
            
            \item \label{item:proof:c} Resampling preserves population sizes and is (conditionally) unbiased, i.e., $R_i=R$ and $\mathbb{E}[r_i^{(j)}]=\tau_i^{(j)}$, where $\tau_i^{(j)}$ originates from Eq.~\eqref{eq:tau}.
            %Population sizes are constant $R_i=R$, while the average number of resampled copies of replica $j$ from $\beta_{i-1}$ to $\beta_i$ is still given by Eq.~\eqref{eq:tau}.

            \item  \label{item:proof:d} No equilibration routines are employed.
        \end{enumerate}
        Moreover, consider $M$ independent simulations of this algorithm employing identical annealing schedules, target distributions, and target population sizes.
        
        Then, the configurational weighted average of the empirical distribution converges almost surely to the target distribution, i.e., it holds with probability one that
        \begin{equation}\label{eq:convergence}
            \lim_{M \to \infty} \W{\widehat{\rho}_i(\gamma)} = \rho_{\beta_i}(\gamma) \qquad \forall \gamma \in \Gamma.
        \end{equation}
        In this case, configurational weighted averages, the weighted free-energy estimator from Sec.~\ref{sec:WA:F} and the central moment estimators introduced below in Sec.~\ref{sec:WA:moments} are asymptotically unbiased.
        This is due to the identity
        \begin{equation}
            \W{\widehat{\mathcal{O}}_i}
            = \sum_{m=1}^M w_i^{(m)} \widehat{\mathcal{O}}_i^{(m)}= \sum_{\gamma \in \Gamma} \W{\widehat{\rho}_i(\gamma)} \mathcal{O}(\beta_i, \gamma)
        \end{equation}
        for every configurational estimator $\widehat{\mathcal{O}}_i$ and the remark at the end of Sec.~\ref{sec:WA:conf}.
        
        \textit{Proof:}
        Due to the assumption of constant and identical population sizes, we have
        \begin{equation}\label{eq:proof:weights}
            w_i^{(m)} = \frac{\prod_{k=1}^i Q_k^{(m)}}{\sum_{m'=1}^M \prod_{k=1}^i Q_{k}^{(m')}}.
        \end{equation}
        Since $\Gamma$ is finite, $\mathbb{E} [\prod_{k=1}^i Q_k]$ exists and Kolmogorov's strong law of larger numbers~\cite[Theorem 11.3.1]{borovkov:13} implies almost sure convergence as $M \to \infty$
        \begin{subequations}
            \begin{align}
                \frac{1}{M} \sum_{m=1}^M \widehat{\rho}_i^{(m)} (\gamma) \prod_{k=1}^i Q_k ^{(m)}&\overset{a.s.}{\longrightarrow}&&\mathbb{E}\left[\widehat{\rho}_i (\gamma) \prod_{k=1}^i Q_k \right],\\
                \frac{1}{M} \sum_{m=1}^M \prod_{k=1}^i Q_{k}^{(m)}&\overset{a.s.}{\longrightarrow}&& \mathbb{E}\left[ \prod_{k=1}^i Q_{k} \right].
            \end{align}
        \end{subequations}
        The existence of these limits guarantees almost surely
        \begin{equation}\label{eq:proof:limit}
            \lim_{M \to \infty} \W{\widehat{\rho}_i(\gamma)} = \frac{\mathbb{E}\left[\widehat{\rho}_i (\gamma) \prod_{k=1}^i Q_k \right]}{\mathbb{E}\left[ \prod_{k=1}^i Q_{k} \right]},
        \end{equation}
        which is a normalized distribution on $\Gamma$ by linearity.
        Hence, it suffices to show that the numerator in Eq.~\eqref{eq:proof:limit} is proportional to $v_i(\gamma)$ up to a constant.
        This is trivial, if we pick $\gamma \in \Gamma$ with $v_i(\gamma)=0$ since resampling at $\beta_{i-1} \mapsto \beta_i$ cannot create replicas in $\gamma$.
        Thus, we may assume $v_i(\gamma)>0$ which, by assumption \ref{item:proof:b}, implies $v_k(\gamma)>0$ for all $k \leq i$.
        Denote the population at $\beta_k$ by $\mathcal{P}_k \in \Gamma^R$ and let
        $\mathcal{P}_0, \ldots, \mathcal{P}_{i-1}$ be any \emph{possible} sequence of populations throughout the anneal.
        A short calculation in App.~\ref{app:proof} using \ref{item:proof:c} and \ref{item:proof:d} yields
        \begin{eqnarray}\label{eq:proof:firstStep}
            \mathbb{E}\Big[ \widehat{\rho}_i(\gamma) \prod_{k=1}^i &Q_k& \Big|\,\mathcal{P}_0, \ldots, \mathcal{P}_{i-1}~\text{fixed}\,\Big]\nonumber \\
            &=& \frac{v_i (\gamma)}{v_{i-1}(\gamma)} \widehat{\rho}_{i-1}(\gamma) \prod_{k=1}^{i-1} Q_k,
        \end{eqnarray}
        where it is also shown that Eq.~\eqref{eq:proof:firstStep} together with the law of total expectation~\cite[Eq. (4.2.2) or p. 98]{borovkov:13} implies
        \begin{align}\label{eq:proof:secndStep}
            \mathbb{E} &\Big[ \widehat{\rho}_i (\gamma) \prod_{k=1}^i Q_k \Big|\,\mathcal{P}_0, \ldots, \mathcal{P}_{i-2}~\text{fixed}\,\Big]\nonumber\\
            &= \frac{v_i(\gamma)}{v_{i-1}(\gamma)}\mathbb{E} \Big[ \widehat{\rho}_{i-1} (\gamma) \prod_{k=1}^{i-1} Q_k \Big|\,\mathcal{P}_0, \ldots, \mathcal{P}_{i-2}~\text{fixed}\,\Big].
        \end{align}
        As demonstrated in App.~\ref{app:proof}, it follows that the recursion in Eq.~\eqref{eq:proof:secndStep} enables one to successively reduce the number of fixed populations until $\beta_0$ is reached
        \begin{eqnarray}
            \mathbb{E} \Big[ \widehat{\rho}_i (\gamma) &\prod_{k=1}^i& Q_k \Big|\, \mathcal{P}_0~\text{fixed}\,\Big]\nonumber\\
            &=& \frac{v_i(\gamma)}{v_1 (\gamma)}\,\mathbb{E} \Big[ \widehat{\rho}_1 (\gamma) Q_1\Big|\, \mathcal{P}_0~\text{fixed}\, \Big].
        \end{eqnarray}
        The right hand side resolves to $[v_i(\gamma)/v_0(\gamma)]\widehat{\rho}_0 (\gamma)$ using Eq.~\eqref{eq:proof:firstStep} at $i=1$.
        Finally, the law of total expectation implies by assumption \ref{item:proof:a}
        \begin{align}
            \mathbb{E}\Big[ \widehat{\rho}_i(\gamma) \prod_{k=1}^i Q_k\Big] &= \frac{v_i(\gamma)}{v_0(\gamma)} \sum_{\mathcal{P}_0 \in \Gamma^R} \mathbb{P}(\mathcal{P}_0) \widehat{\rho}_0 (\gamma)\\
            &= \frac{v_i(\gamma)}{v_0(\gamma)} \mathbb{E}[\widehat{\rho}_0(\gamma)]\\
            &= v_i (\gamma) / C_0, 
        \end{align}
        which completes the proof and also shows that the numerator in Eq.~\eqref{eq:proof:limit} equals $C_i/C_0$.
        
        We anticipate a similar statement to hold in the presence of appropriate equilibration routines as they serve to reduce systematic errors in each individual run already.
        However, a rigorous proof in this setting would presumably require a more advanced mathematical treatment.
        Apart from restriction \ref{item:proof:b}, which is only needed due to \ref{item:proof:d},
        no additional constraints on the annealing schedule or target distributions are necessary other than the remarks in Secs.~\ref{sec:ALG:ideas} and~\ref{sec:ALG:alg} required to run PA in the first place.

    \subsection{Central moments}\label{sec:WA:moments}
    
        We now want to discuss further important examples of estimators that are not configurational. An important class of such quantities are (empirical) central moments of configurational estimators,
        \begin{equation}\label{eq:k-th_moment}
            \widehat{\mathcal{K}}(\beta) \coloneqq \int_{\Gamma} \Big[\mathcal{O}(\beta, \gamma) - \widehat{\mathcal{O}} \Big]^k \, \widehat{\rho}_{\beta}(\gamma)\, \text{d}\gamma,\quad k \in \mathbb{N},
        \end{equation}
        which most importantly includes sample variances.
        Throughout this section, we omit indices related to the annealing schedule and use subscripts to indicate the order of moments.
        To this end, let $\mu_l$ be the $l$-th central moment of some random variable and $\mu'_l$ be the $l$-th moment about the origin, provided they exist.
        It follows from the binomial theorem that $\mu_k$ is determined by $\mu_1', \ldots, \mu_k'$ via
        \begin{equation}
            \mu_k = \sum_{l=0}^k \binom{k}{l}(-1)^{k-l} (\mu_1')^{k-l} \mu_l',
        \end{equation}
        where $\mu_0' = 1$.
        Applying this to Eq.~\eqref{eq:k-th_moment}, we can express $\widehat{\mathcal{K}}$ in terms of ensemble averages of $\mathcal{O}, \ldots, \mathcal{O}^k$, which yields the PA estimator
        \begin{equation}\label{eq:k-th_moment_est}
            \widehat{\mathcal{K}} \coloneqq \sum_{l=0}^k \binom{k}{l} (-1)^{k-l} \big(\widehat{\mathcal{O}}\big)^{k-l} \widehat{\mathcal{O}^{l}},
        \end{equation}
        where $\widehat{\mathcal{O}}, \ldots, \widehat{\mathcal{O}^k}$ are the population averages of the respective power of $\mathcal{O}$ according to Eq.~\eqref{eq:PopAvrg}.
        Note that Eq.~\eqref{eq:k-th_moment_est} defines a continuous function of configurational estimators from which the appropriate weighted estimator for $\mathcal{K}$ is obtained
        \begin{equation}
            \mathcal{W}_{k} \big[\widehat{\mathcal{K}}\big] \coloneqq \sum_{l=0}^k \binom{k}{l} (-1)^{k-l} \big(\W{\widehat{\mathcal{O}}}\big)^{k-l} \W{\widehat{\mathcal{O}^{l}}}.
        \end{equation}
        Almost sure convergence of configurational weighted averages thus directly implies that weighted estimators for arbitrary central moments are asymptotically unbiased.
        
        Since we will mainly focus on the case $k=2$, the more instructive notation $\mathcal{W}_{\text{var}}$ is used to denote the \emph{weighted variance estimator}, which is compared to the falsely applied configurational weighted average $\W{\widehat{\mathcal{K}}}$ in Sec.~\ref{sec:num:bias} and~\ref{sec:num:stat}.
        In particular, we consider the weighted heat capacity estimator,
        \begin{equation}\label{eq:Wvar_C}
             \Wvar{\widehat{c}} \coloneqq \beta^2 N \left[ \W{\widehat{e^2}} -  \Big(\W{\widehat{e}}\Big)^2 \right],
        \end{equation}
        and also the susceptibility of the Ising FM,
        \begin{equation}\label{eq:Wvar_Chi}
             \Wvar{\widehat{\chi}} \coloneqq \beta N \left[ \W{\widehat{m^2}} -  \Big(\W{\widehat{m}}\Big)^2 \right],
        \end{equation}
        which may be compared to Eqs.~\eqref{eq:c} and~\eqref{eq:chi}.
        The weighted variance estimator is bounded from below by the falsely applied configurational weighted average
        \begin{equation}\label{eq:Wvar_bound}
            \Wvar{\widehat{\mathcal{K}}} = \W{\widehat{\mathcal{K}}} + \sum_{m=1}^M w^{(m)} \Big(\widehat{\mathcal{O}}^{(m)} - \W{\widehat{\mathcal{O}}}\Big)^2.
        \end{equation}
        This is due to the fact that the configurational weighted average of the sample variance does not take into account fluctuations of the sample mean between independent simulations and thus underestimates the actual variance.
        For a numerical demonstration, see Fig.~\ref{fig:c_chi_bias} at $M=50$.
        %This form is plausible as latter does not take into account the additional fluctuation from variation of the means between the different PA runs.
    
    \subsection{Spin overlap}\label{sec:WA:q}
    
        Wang \emph{et al.}~\cite{wang:15a} not only proposed a way to measure spin overlaps in one PA simulation, but also claimed that configurational weighted averaging works for the spin overlap distribution, i.e., they introduced the estimator~\cite{wang:15a}
        \begin{equation}\label{eq:W_conf_q}
            \W{\widehat{P}_\mathcal{J}(q)} \coloneqq \sum_{m=1}^M w^{(m)} \widehat{P}_\mathcal{J}^{(m)}(q),
        \end{equation}
        where $\widehat{P}_\mathcal{J}^{(1)}(q), \ldots, \widehat{P}_\mathcal{J}^{(M)}(q)$ are empirical distributions obtained from independent runs.
        In contrast to Wang~\emph{et al.}, we apply this formula to data $\widehat{P}_\mathcal{J}(q)$ measured by the index shift approach, due to its better parallel efficiency among other reasons discussed in Sec.~\ref{sec:MO:Overlap}.
        The average of Eq.~\eqref{eq:W_conf_q} over several disorder realizations is used as an estimator for $P(q)$,
        \begin{equation}\label{eq:W_conf_q_disAvrg}
            \W{\widehat{P}(q)} \coloneqq \left[\W{\widehat{P}_\mathcal{J}(q)}\right]_\mathrm{av}.
        \end{equation}
        Even more than in case of the single-run measurement $\widehat{P}_\mathcal{J}(q)$, the stability of the weighted estimator $\W{\widehat{P}_\mathcal{J}(q)}$ depends strongly on the number of surviving families.
        If replicas are poorly equilibrated, the occasional encounter of low-energy states results in a massive decline in surviving families due to the rapid reproduction of such configurations.
        Consequently, $q$ values from this run are correlated since members of the largest family are included in a significant fraction of pairings.
        At the same time, the presence of relative low-energetic states implies larger free-energy weights, thereby potentially attaching a high weight to a PA simulation of already weak family statistics.
        
    \subsection{Variance of free-energy weights}\label{sec:WA:w}
    
        Lastly, we show that the variance of free-energy weights can be predicted rather accurately for sufficiently large population sizes. 
        It follows from the central limit theorem that the free-energy estimator $\widehat{F}$ is normally distributed in the limit $R \to \infty$~\cite{wang:15a,weigel:21}.
        Given that simulations of identical target population size are considered, we may disregard population size related terms in Eq.~\eqref{eq:appropriateWeight} and \eqref{eq:simpleWeight} whose effect seems to be rather small numerically.
        Thus, we arrive at
        \begin{equation}
           w_i^{(m)} = \frac{\exp(-\beta_i \widehat{F}_i^{(m)})}{\sum_{m'=1}^M \exp(-\beta_i \widehat{F}_i^{(m')})} = \underline{w}_i^{(m)}.
        \end{equation}
        If additionally $M$ is large or the distribution of $\widehat{F}$ narrow, $w_i^{(m)}$ is the exponential of a Gaussian variable with an approximately ``constant'' prefactor scaling its mean to $1/M$.
        Thus, $w_i^{(m)}$ is log-normal in this limit~\cite{weigel:21}.
        Since $\widehat{F}_i^{(1)}, \ldots, \widehat{F}_i ^{(m)}$ are i.i.d., it follows from mean and variance of log-normal variables that
        \begin{equation}\label{eq:varWpred}
            \text{var}\, w \approx \frac{\text{var}\, \exp (- \beta \widehat{F})}{(M \, \mathbb{E}\exp (- \beta \widehat{F}) )^2} =  \frac{\exp(\text{var}\, \beta \widehat{F}) -1 }{M^2}.
        \end{equation}
        This formula is tested numerically in Sec.~\ref{sec:num:dist}.
        Note, however, that the right-hand side of Eq.~\eqref{eq:varWpred} is unbounded while, in contrast, the actual variance trivially cannot exceed one.
        
\section{Numerical results}\label{sec:num}

    We now turn to a detailed comparison of the theoretical concepts for weighted averages discussed above with an extensive array of PA simulations for the two-dimensional Ising FM and SG.
    A description of our methodology is given in Sec.~\ref{sec:Metho} including details of our implementation, our simulation data, the way in which it was processed and the reference solutions needed to calculate systematic errors. 
    Moreover, our notion of ``difficult'' disorder realizations is explained.

    The presentation of numerical results itself has a tri-fold structure starting with the most important aspect of bias reduction through weighted averaging in Sec.~\ref{sec:num:bias}.
    Particular emphasis is placed on the weighted variance estimator, the spin overlap distribution and the exponent of a potential power-law decay of bias with respect to $M$.
    Secondly, we investigate previous claims \cite{wang:15a,weigel:21} regarding the distribution of the free-energy estimator $\widehat{F}$ in Sec.~\ref{sec:num:dist} before addressing the question to which extent bias is reduced at the cost of larger statistical errors in Sec.~\ref{sec:num:stat}.

    \subsection{Methodology}\label{sec:Metho}

        \subsubsection{Implementation}\label{sec:Metho:Impl}
        
            Our simulations of the Ising FM employ the optimized GPU implementation provided by Barash~\emph{et~al.}~\cite{barash:16}.
            Only slight modifications are needed to adapt the code to the Ising SG such that essentially the same program was used for both models.
            Unless mentioned otherwise, spin overlap measurements were conducted by choosing pairs via index shifts as explained in Sec.~\ref{sec:MO:Overlap}.
            
            During the resampling step \ref{alg:step:iii} of the algorithm given in Sec.~\ref{sec:ALG:alg}, copies of the same ancestor are placed next to each other in replica index space~\cite{barash:16} which localizes correlations and allows to measure the performance of the equilibration routine~\cite{weigel:21}.
            Step \ref{alg:step:iv} consists of single-spin-flip Metropolis updates, and $\theta$ sweeps are performed at every temperature.
            Additionally, a checkerboard decomposition allows to modify spins inside the same sublattice in parallel, see Ref.~\cite{barash:16} for further details.
            The resulting update scheme does not satisfy detailed balance, but meets the required global balance condition~\cite{potter:2013}.
            
        \subsubsection{Conducted simulations and averaging}\label{sec:Metho:Sim}
        
            We applied an equidistant annealing schedule of inverse temperatures $\beta_i \coloneqq i \Delta \beta, i \geq 0$ using $\Delta \beta=0.005$ and $\Delta \beta = 0.03$ for Ising FM and SG, respectively, terminating at $\beta_f=1$ (FM) and $\beta_f=3$ (SG).
            The target population size was chosen to be $R=2 \times 10^4$ for all simulations (apart from the reference runs described in Sec.~\ref{sec:Metho:Sol}).
            This value should be substantially larger in PA simulations aiming to study unknown systems reliably~\cite{weigel:21},
            but in contrast here we are interested in exposing systematic errors.
            To this end, the number of Metropolis sweeps and the target population size are picked rather small on purpose to more clearly see the resulting artifacts.
            %This not only implies picking rather small numbers of Metropolis sweeps on purpose, but also using smaller population sizes to measure resulting artifacts.
            
            For the same reason, we solely investigated 50 randomly generated $L=32$ disorder instances and ran numerous repeated simulations to effectively eliminate statistical errors:
            $5 \times 10^4$ runs were conducted for each instance and $\theta \in \{2,5,10\}$ was chosen to take into account different equilibration levels.
            In combination with the reference runs described in Sec.~\ref{sec:Metho:Sol} and the additional simulations for Fig.~\ref{fig:qHist_WangComparison} this resulted in a total of more than $7.7 \times 10^6$ independent SG simulations.
            
            For given $M$, $\theta$ and a specific realization, the pool of independent runs was randomly partitioned into $S=\floor{5 \times 10 ^4 / M}$ subgroups within which weighted averaging over $M$ simulations was performed.
            Finally, we trivially averaged over these $S$ samples of weighted estimators to measure their mean values.
            Since resulting bias estimates for the same disorder instance, but different values of $M$, share the same pool of simulations, they may be slightly correlated.
            In view of having $5 \times 10^4$ runs to choose from, we disregard such effects, however.
            
            Simulations for the Ising FM include different values of $\theta$ and system sizes $L \in \{16,32,64,128\}$, although the majority of our data was collected at $L=64$.
            As this model is computationally less expensive, we used entirely independent simulations for different values of $M$.
            That is, for fixed $L, \theta$ and $M$ we obtained $S$ samples of weighted estimators, each consisting of $M$ separate PA runs which are independent to all other runs including those for different $M$.
            For every choice of simulation parameters, we ensured that $S \geq 5\times 10^3$ while using $S=8 \times 10^3$ for $M \leq 15$.
            In total, data from at least $3.1 \times 10^6$ individual Ising FM simulations are shown in the figures.

        \subsubsection{Reference solutions}\label{sec:Metho:Sol}
        
            The examples of the two-dimensional Ising FM and SG introduced in Sec.~\ref{sec:MO:Models} were chosen to allow for precise bias measurements by merit of the available  exact solutions.
            Onsager famously solved the ferromagnetic model in the limit $L \to \infty$~\cite{onsager:1944} and for finite $L$ explicit results for $Z(\beta)$ are available as well~\cite{kaufman:1949, fisher:1969}.
            The two-dimensional Ising SG admits the evaluation of $Z_{\mathcal{J}}(\beta)$ for a given disorder realization $\mathcal{J}$ and inverse temperature~$\beta$ by efficient algorithms such as the publicly available implementation by Thomas and Middleton~\cite{thomas:13} which has time complexity $\mathcal{O}(L^3)$.
            Thus, we were able to evaluate the partition function of both models to obtain exact values for the internal energy, heat capacity and free energy.
            
            In contrast, we are unaware of efficient methods to calculate the susceptibility $\chi$ of the Ising FM or observables related to the spin overlap $q$.
            We therefore reverted to quasi-exact solutions, i.e., measurements from particularly large and well equilibrated PA simulations, which were treated as being exact to enable bias estimations.
            The reference values were obtained by arithmetic averaging over multiple runs with parameters shown in Table~\ref{tab:RefParams}.
            For the SG problem, we performed 100 reference runs for the same disorder instance to drive down statistical errors and partitioned them into 50 pairs to compute approximately $R$ spin overlap values between two runs forming a pair.
            Hence, the reference $q$ distribution for each spin glass instance originates from approximately $50\times R=2.5 \times 10^7$ measured $q$-values.
            The smaller value of $\theta=25$ compared to the FM case was chosen since we did not observe any change in the histograms $\widehat{P}_\mathcal{J}(q)$ on further increasing the number of Metropolis sweeps.

            \begin{table}[b]
                \centering
                \caption{\label{tab:RefParams}
                Parameters used for quasi-exact reference PA simulations for the Ising FM and Ising SG instances. Shown are linear system size $L$, target population size $R$, the number of Metropolis sweeps $\theta$, independent repetitions and the number of considered disorder instances $\mathcal{J}$.}
                \begin{ruledtabular}
                \begin{tabular}{cccccc}
                    \text{system} & $L$ & $R$ & $\theta$ & \text{runs} & \text{sampled}~$\mathcal{J}$\\
                    \colrule
                    FM & $16$ & $10^6$ & $50$ & $2 \times 10^4$ & - \\
                    FM & $32$ & $10^6$ & $50$ & $10^4$ & - \\
                    FM & $64$ & $10^6$ & $70$ & $10^4$ & - \\
                    FM & $128$ & $10^6$ & $80$ & $10^4$ & - \\
                    SG & $32$ & $5 \times 10^6$ & $25$ & $100$ & $50$ \\
                \end{tabular}
                \end{ruledtabular}
            \end{table}
            
        \subsubsection{Hardness of realizations}\label{sec:Meth:rhot}
        
            Although we only considered a small number of 50 Ising SG instances, this is sufficient to illustrate the sometimes variable results of weighted averaging depending on the ``hardness'' of disorder realizations.
            Due to the availability of various PA equilibration metrics~\cite{machta:10a,wang:15a, weigel:21}, such instances can be conveniently identified.
            Here, we relied on the \emph{mean square family size}~\cite{wang:15a}
            \begin{equation}\label{eq:rho_t}
                \rho_t(\beta_i) \coloneqq R_i \sum_{k} \mathfrak{n}_{k,i}^2,
            \end{equation}
            where $\mathfrak{n}_{k,i}$ is the fraction of replicas at $\beta_i$ in family $k$.
            Large $\rho_t$ indicates the existence of large families, thereby often resulting in poor sampling quality.
            Throughout this paper, references such as ``hardest'' instance refer to comparisons of the mean value $\overline{\rho_t}$ for fixed PA parameters and inverse temperatures.

    \subsection{Bias reduction}\label{sec:num:bias}
    
        \subsubsection{General behavior}
    
            \begin{figure}
                \centering
                \includegraphics[width=0.95\linewidth]{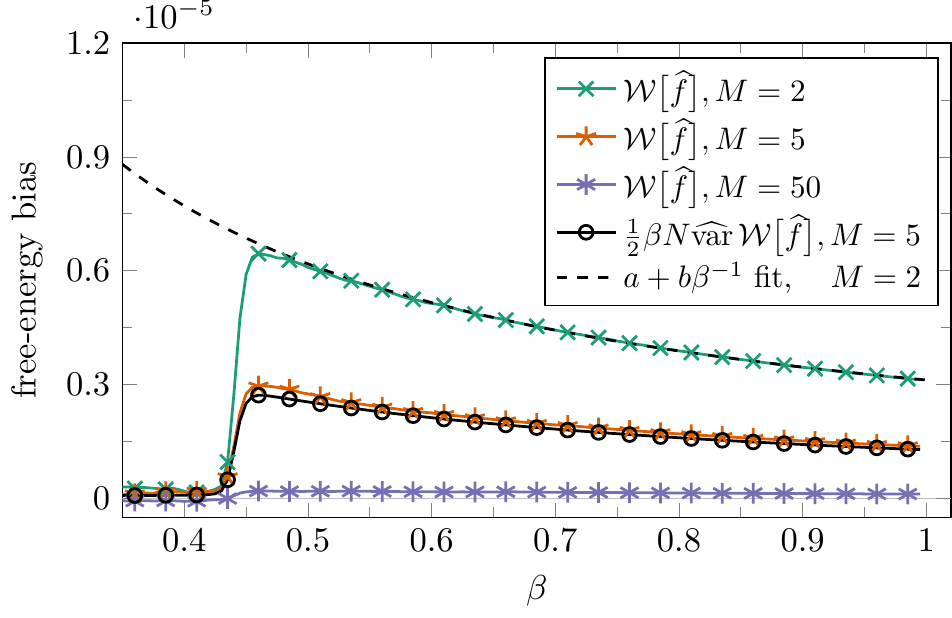}
                \caption{Measured bias of the weighted free-energy per spin estimator $\W{\widehat{f}}$ of the $L=64$ Ising FM at $\theta=10$. The estimate $\frac{\beta}{2}\text{var}\, \widehat{F}$ was proposed by Wang \emph{et al.} \cite{wang:15a} in the limit of large populations and was adjusted here for the weighted average of $f$. The data for $\beta \geq 0.6$ were used to determine the fit to the $M=2$ curve drawn as the dashed line.
                Only every fifth data point is highlighted on each of the curves.}
                \label{fig:F_bias}
            \end{figure}
            
            As a first example, Fig.~\ref{fig:F_bias} shows systematic errors of the weighted free-energy estimator $\W{\widehat{F}}$ applied to the Ising FM.
            Near the critical (inverse) temperature $\beta_c =\tfrac{1}{2}\ln(1+\sqrt{2}) \approx 0.4407$, populations start deviating from the equilibrium distribution due to critical slowing down, resulting in a steep bias increase.
            This can be compensated by weighted averaging, however, such that the bias steadily decreases for an increasing number of runs, and systematic errors are no longer discernible compared to the statistical errors for $M=50$ simulations, cf.\ Fig.~\ref{fig:F_bias}.
            Moreover, the reduction is mostly uniform with respect to $\beta$, rendering this the prototypical situation of successful weighted averaging.
            Wang \emph{et al.}~\cite{wang:15a} suggested that the bias of the (non-weighted) free-energy estimator $\widehat{F}$ is given by $\frac{\beta}{2}\text{var}\, \widehat{F}$ for large population sizes $R$. They also conjectured that the same formula should be a good approximation for the  weighted estimator, i.e., when replacing $\widehat{F}$ by $\W{\widehat{F}}$. This indeed works well here, as is illustrated by the corresponding data in Fig.~\ref{fig:F_bias}.
            If the same formula is applied to less well equilibrated runs, the difference between actual bias and the prediction can be significantly larger, however (see also Ref.~\cite{weigel:21}).
            Finally, the dashed curve represents the least-squares fit of $a + b \beta^{-1}$ to the $M=2$ data for $\beta \geq 0.6$.
            Recall that we argued in Sec.~\ref{sec:MO:F} and App.~\ref{app:F_bias} that systematic errors of $\widehat{F}$ in the Ising FM asymptotically behave as $\beta^{-1}$; this apparently generalizes to $\W{\widehat{F}}$.
            
            \begin{figure}
                \centering
                \includegraphics[width=\linewidth]{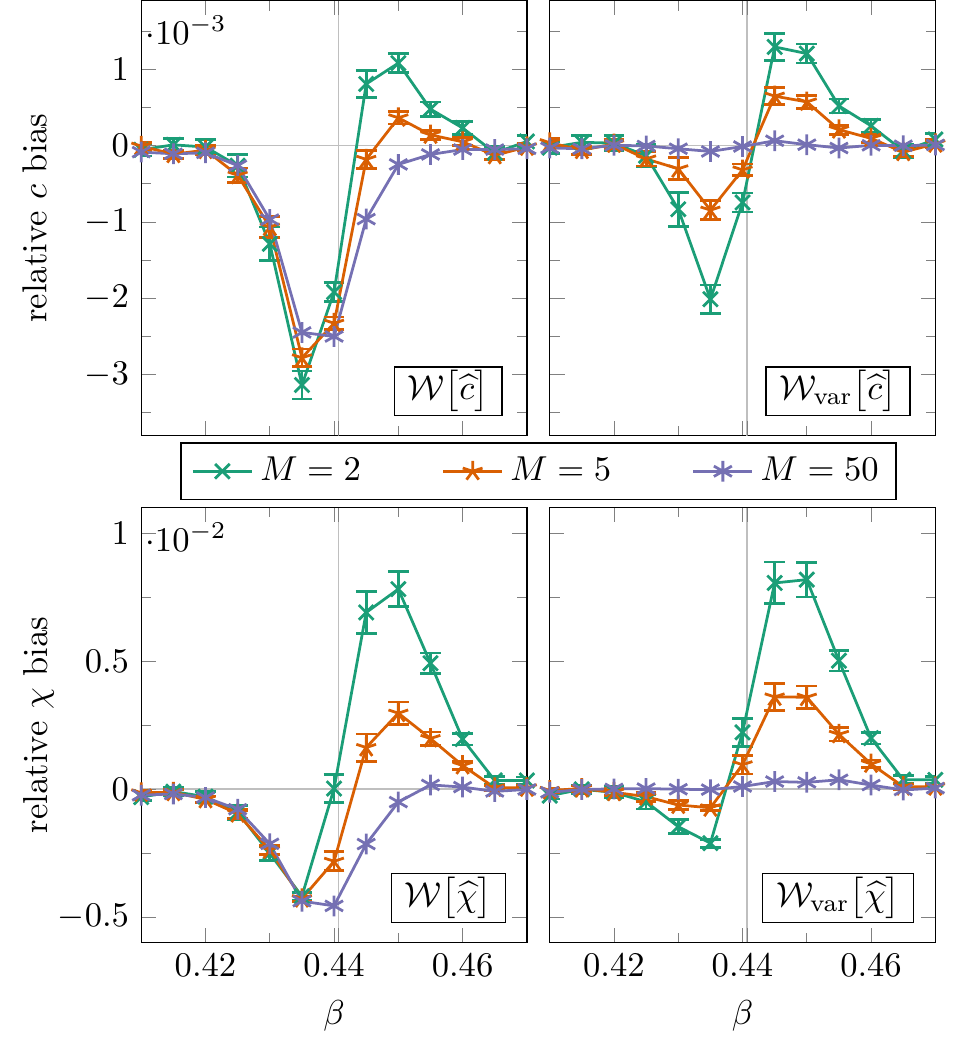}
                \caption{Bias comparison in measuring the heat capacity $c$ and susceptibility $\chi$ of the Ising FM through configurational weighted averages $\mathcal{W}$ (left) or the weighted variance estimator $\mathcal{W}_\text{var}$ (right).
                The system size is $L=64$, $\theta=10$ Metropolis sweeps were used, and $M$ represents the number of independent simulations entering weighted averaging.
                Error bars at $M=50$ are significantly smaller than the symbols.
                Relative bias is used since $c$ and $\chi$ vary strongly in the vicinity of $\beta_c \approx 0.4407$ (marked by the vertical line).}
                \label{fig:c_chi_bias}
            \end{figure}
            
            Next, we would like to point out the importance of choosing the appropriate weighted estimator using the example of the heat capacity and susceptibility. 
            Note that the estimators $\widehat{c}$ and $\widehat{\chi}$ from Eqs.~\eqref{eq:c} and~ \eqref{eq:chi} are not configurational as they cannot be expressed in terms of a single ensemble average [see Eq.~\eqref{eq:configurational}] unless the respective mean values are known \emph{a priori}.
            % Note that these quantities are not configurational since $c$ and $\chi$ cannot be expressed in terms of a single ensemble average [see Eq.~\eqref{eq:configurational}] unless the respective mean value is known \emph{a priori}.
            The systematic errors that result when the (wrong) configurational weighted average $\mathcal{W}$ is applied are depicted in the left panels of Fig.~\ref{fig:c_chi_bias}, whereas Eqs.~\eqref{eq:Wvar_C} and~\eqref{eq:Wvar_Chi} were employed on the right hand side.
            In the latter case, bias is reduced uniformly at all temperatures as more simulations are taken into account.
            In contrast, falsely applying configurational weighted averages results in dominant systematic errors, which may even surpass those of single PA runs.
            In view of Eq.~\eqref{eq:Wvar_bound}, one expects negative bias for $\mathcal{W}$ since it misses a non-negative term that contains contributions due to the variations of the population averages $\widehat{e}$ and $\widehat{m}$.
            
            Turning to the simulations of the SG system, we find that for single disorder realizations weighted estimators of the energy, free energy and heat capacity behave fairly similar to the results shown for the Ising FM in Figs.~\ref{fig:F_bias} and \ref{fig:c_chi_bias}.
            Although bias curves occasionally display more complex behavior, the overall trend remains that systematic errors of correctly weighted estimators are uniformly reduced in $\beta$ by increasing $M$.
            
            \begin{figure}
                \centering
                \includegraphics[width=\linewidth]{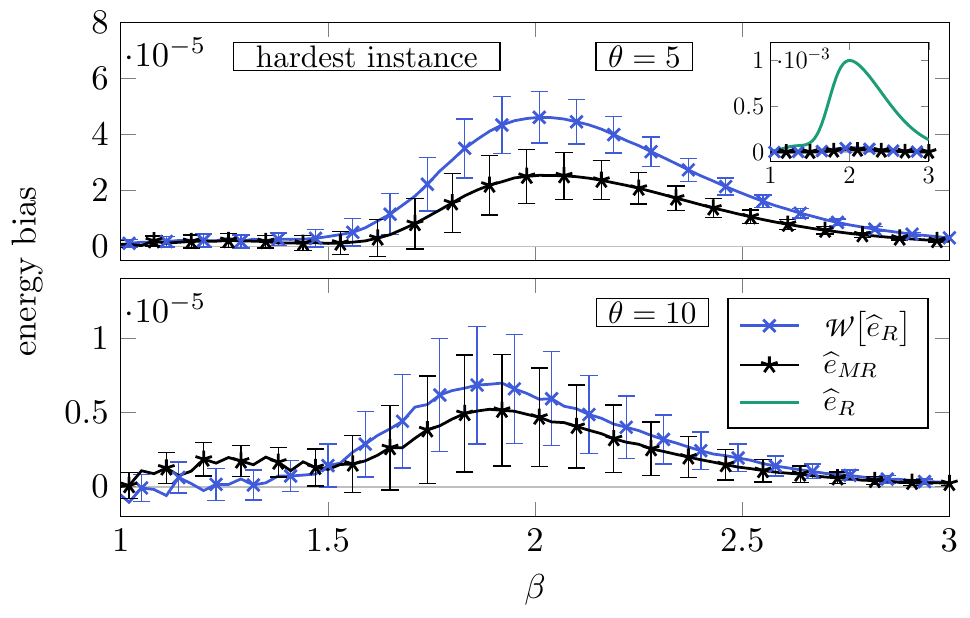}
                \caption{\label{fig:SG_bias_MR}
                Energy bias comparison between weighted averaging over $M=50$ runs of size $R$ and increasing the population size to $MR$ for the ``hardest'' $L=32$ Ising SG instance at $\beta=2.4$, using $\theta=5$ (upper panel) and $\theta=10$ (lower panel), respectively.
                For improved readability, symbols and statistical errors are only drawn at every third annealing step in the main plots and at every tenth step in the $\theta=5$ inset.
                Error bars show the standard deviation of the mean based on $10^3$ repetitions for both approaches.
                Thus, they can be used to compare the standard deviation of the actual estimators.
                Bias of single PA runs at size $R$ is substantially larger, which is illustrated by the upper curve in the inset.
                }
            \end{figure}
            
            An important natural benchmark in employing weighted averages is the comparison of $M$ runs of size $R$ with a single run of population size $MR$.
            Due to limited computational resources, we only conducted two such comparisons, one for the ``hardest'' and one for the ``easiest'' SG instance (see Sec.~\ref{sec:Meth:rhot}) at $\beta=2.4$, $M=50$ and the equilibration levels $\theta\in\{5,10\}$.
            Additional to the $S=5\times 10^4 / 50 = 10^3$ weighted estimators, we ran $10^3$ repeated simulations with population size $MR$.
            Similarly to the analogous curves for $c$ and $f$, the resulting energy estimations are remarkably accurate which is shown in Fig.~\ref{fig:SG_bias_MR} for the ``hardest'' instance.
            Although there is a clear difference in bias at $\theta=5$, note that these signals only become significant after $10^3$ repetitions and the great majority of systematic errors is successfully reduced, as can be seen in the upper panel inset showing the plain average of the runs of size $R$ for comparison.
            Statistical errors are also comparable, which can be inferred from the error bars as explained in the caption.
            Hence, it is unlikely that one is able to reliably tell both estimators apart based on a smaller data set, although one would not consider simulations of this instance to be in equilibrium at $\theta=5$ (for which $\overline{\rho_t}\approx 0.4 \times R$ at $\beta=2.4$) and only moderately well equilibrated at $\theta=10$.
            For the ``easiest'' instance at $\beta=2.4$, weighted averaging and the scaled population size are virtually indistinguishable when measuring energy, heat capacity or free energy using $\theta \in \{5,10\}$ Metropolis sweeps (not shown).
            A similar comparison was also conducted for the $L=64$ Ising FM at $\theta=10$ and $M\in \{30,50\}$, resulting in very similar conclusions for measurements of $e,c,f$ and $\chi$.

        \subsubsection{Weighted spin overlap measurements}
    
            \begin{figure}
                \centering
                \includegraphics[width=\linewidth]{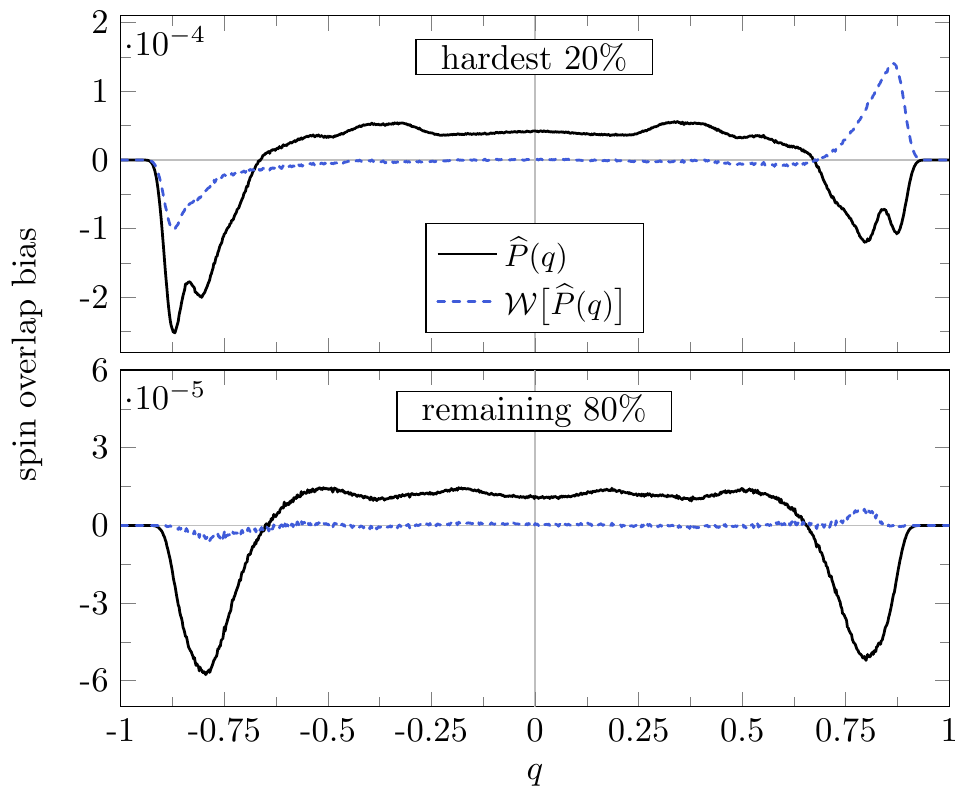}
                \caption{\label{fig:bias_qhist}Systematic errors in spin overlap measurements of $L=32$ Ising SG instances at $\beta=2.4$ for single PA runs (solid line) and weighted averages (dashed line).
                The data sets are averaged over instances of the respective difficulty as explained in the main text.
                Although only a fraction of realizations is taken into account in each case, we adapt the notation from Eq.~\eqref{eq:W_conf_q_disAvrg} here.
                In both cases, $\theta=10$ equilibration sweeps were employed and $M=50$ runs are used to form the weighted average.
                Error bars (not including sample-to-sample contributions) were omitted for clarity as they are negligible in the upper panel and at the magnitude of visible fluctuations below.}
            \end{figure}
            
            For the spin-glass problem, it is well known that the overlap is slower to equilibrate than the energy (see, e.g., Ref.~\cite{houdayer:01}). Additionally, it is more difficult to get a reliable estimate of the whole distribution $P(q)$ than for a single moment.
            Thus, it comes as no surprise that the measured histograms are noticeably asymmetric at $\theta \in \{2,5\}$ and $\beta > 1$, which violates the spin-flip symmetry of the Hamiltonian \eqref{eq:H_SG}.
            We conclude that these equilibration levels and the population size $R=2 \times 10 ^4$ are insufficient for reliable $q$ measurements and therefore focus on $\theta=10$.
            
            This ensures decent equilibration for most disorder instances while a small proportion is still far enough from equilibrium to infer prototypical behavior for such cases.
            To give an example, we consider the ``hardest 20\%'' of disorder realizations at $\beta=2.4$ in the sense of Sec.~\ref{sec:Meth:rhot}.
            The $\rho_t$ threshold we obtain in this way is $1997 \approx R/10$.
            Averaging of the measured bias values at $\beta=2.4$ over the instances grouped in this fashion, we arrive at the result shown in Fig.~\ref{fig:bias_qhist}.
            
            Weighted averaging based on $M=50$ independent runs applied to the ``hardest 20\%'' significantly increases bias at large values of $q \approx 0.8$, cf.\ the upper panel of Fig.~\ref{fig:bias_qhist}.
            The $\pm q$ asymmetry in the arithmetic average is further amplified by weighted averaging, visibly worsening the measurement due to the lack of diversity among surviving families.
            However, the procedure works reasonably well for the ``remaining 80\%'', compensating negative bias for large absolute spin overlaps as is visible in the lower panel of Fig.~\ref{fig:bias_qhist}.
            Still, there is a slight but significant overcompensation at $q \approx 0.8$ which is reminiscent of ``harder'' instances.
            A similar, yet amplified, behavior as in the upper panel of Fig.~\ref{fig:bias_qhist} is observed at $\theta\in\{2,5\}$ for the majority of instances.

            \begin{figure}
                \centering
                \includegraphics[width=\linewidth]{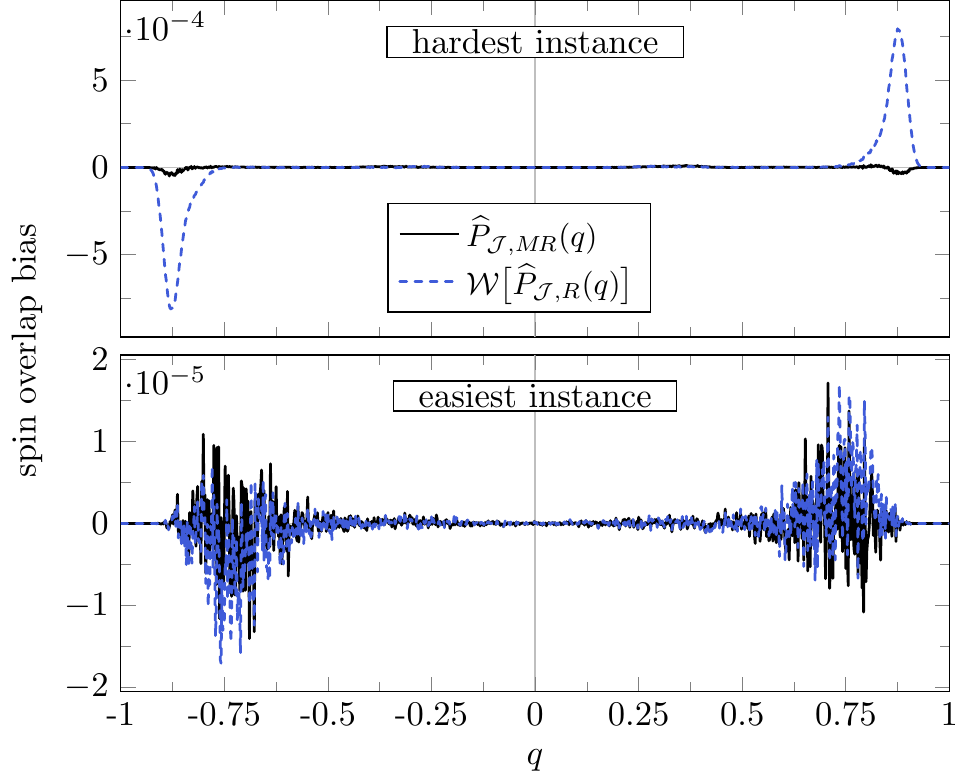}
                \caption{\label{fig:bias_qHist_MR}Bias in measuring the spin overlap distribution of $L=32$ Ising SG instances. Employing weighted averaging (dashed line) to $M=50$ independent runs of population size $R=2 \times 10^4 $ is compared to running single PA simulations (solid line) with population size $MR=10^6$.
                Both the ``hardest'' and ``easiest'' instance encountered at $\beta=2.4$ use $\theta=10$ Metropolis sweeps throughout the annealing process.
                The number of conducted runs of size $MR$ is $10^3$.
                Statistical errors (not including sample-to-sample contributions) are negligible in the upper panel and on the scale of the visible fluctuations below.
                }
            \end{figure}
            
            Nevertheless, it is possible for weighted averaging over $M$ runs of size $R$ to reach the quality of single simulations of size $MR$, at least for particularly ``easy'' instances.
            This is illustrated in Fig.~\ref{fig:bias_qHist_MR}.
            For the ``easiest'' instance, a difference in systematic errors between weighted averaging over $M=50$ runs and scaling the population size by the same factor is barely measurable, even after thousands of repetitions.
            For the ``hardest'' instance, however, $q$ measurements from the same simulation are correlated since descendants from large families are present in virtually every replica pair, resulting in the same artifacts as in Fig.~\ref{fig:bias_qhist}.
            Here, larger population sizes are desperately needed to avoid such behavior and cannot be replaced by weighted averaging since it is unable to remove correlation within simulations.
            If we compare the upper panel of Fig.~\ref{fig:bias_qHist_MR} to the lower panel of Fig.~\ref{fig:SG_bias_MR} where data from the same simulations are shown, we see even more clearly that $\theta=10$ is in principle not insufficient for moderate equilibration.
            Thus, the actual bottleneck for $q$ measurements of this realization is that $\theta$ is small enough for families to regularly reach sizes no longer manageable at $R=2 \times 10 ^4$.
            To underline this, we may additionally compare Figs.~\ref{fig:qHist_WangComparison} and \ref{fig:bias_qHist_MR}.
            The $\pm q$ symmetry in the former depiction is not in contradiction to the lack of symmetry in the latter, since simulations with large families (and therefore correlated $q$ measurements) are removed in Fig.~\ref{fig:qHist_WangComparison} to allow for the intended comparison (see also Table~\ref{tab:qHist_Term}), which is sufficient to restore the symmetry between $\pm q$ at the same value of $\theta=10$.
            It also follows from this comparison that, although the resulting histogram of the ``hardest'' instance in Fig.~\ref{fig:bias_qHist_MR} is dominated by correlation artifacts, it still outperforms the measurements without weighted averaging from Fig.~\ref{fig:qHist_WangComparison}.

            \begin{figure}
                \centering
                \includegraphics[width=\linewidth]{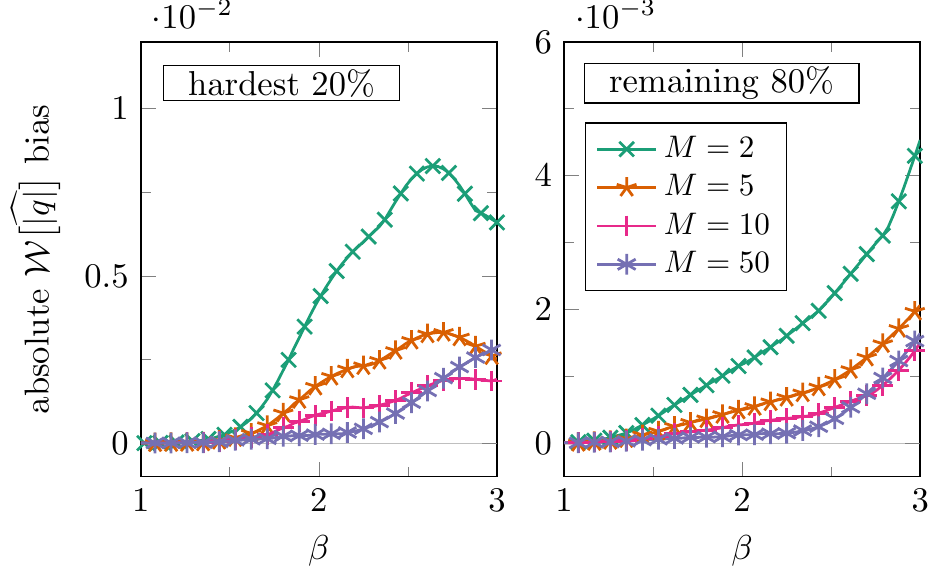}
                \caption{\label{fig:bias:SG_absQ} 
                Disorder average of the absolute bias of configurational weighted averages applied to $\widehat{|q|}$, i.e., the mean absolute value of $q$ measurements.
                Disorder instances of the $L=32$ Ising SG were grouped based on their ``hardness'' at $\beta=2.4$, as explained in the main text.
                $\theta=10$ equilibration sweeps were used in both panels.
                Statistical errors (not including sample-to-sample contributions) are significantly smaller than the symbols which are only drawn at every third data point.
                }
            \end{figure}
            
            We now turn to the spin glass order parameter, i.e., the ensemble average of $|q|$ as discussed in Eq.~\eqref{eq:ensembleAvrg_absQ}.
            If we denote the mean absolute value of all $q$ measurements obtained within a simulation by $\widehat{|q|}$, we apply the configurational weighted average $\W{\widehat{|q|}}$.
            This is equivalent to estimating the expected absolute value based on $\W{\widehat{P}(q)}$.
            Consequently, one may hope that bumps as in the upper panels of Figs.~\ref{fig:bias_qhist} and \ref{fig:bias_qHist_MR} at $q \approx \pm 0.8$ are sufficiently anti-symmetric to cancel when calculating $\W{\widehat{|q|}}$.
            
            In order to probe the bias reduction through $\W{\widehat{|q|}}$, we averaged the absolute systematic error within the ``hardest 20\%'' and ``remaining 80\%'' of disorder instances, which results in Fig.~\ref{fig:bias:SG_absQ}.
            Although the difficulty of instances is temperature-dependent, we presume the groups, originally formed at $\beta=2.4$, to be a reasonable approximation.
            For $\theta=10$, systematic errors decrease through weighted averaging over the whole temperature range while the factor of bias reduction is way below $M$ and visibly worsens at lower temperatures.
            Most strikingly, it is not even monotonic with respect to $M$ in contrast to the case of the observables in Fig.~\ref{fig:F_bias} and \ref{fig:c_chi_bias}.
            Again, this is due to the insufficiency of $R=2 \times 10^4$ at low temperatures and the resulting correlations being amplified by weighted averaging, thereby altering the otherwise monotonous $M$-dependence.
            
            Hence, there is a crucial difference between $q$ measurements and observables which are not defined on $\Gamma \times \Gamma$ such as $e$, $c$ and $f$.
            While our data indicates that appropriate weighted estimators of the latter class reduce systematic errors even for simulations far from equilibrium, this cannot be said in full generality for the spin overlap.
            Insufficient equilibration will lead to correlated $q$ measurements within the same simulation whenever $R$ is too small.
            In the worst case, such correlations are even amplified by free-energy weights such as in the upper panels of Fig.~\ref{fig:bias_qhist} and~\ref{fig:bias_qHist_MR}.
            One should therefore \textit{carefully monitor equilibration in conjunction with population size before applying weighted averages to spin overlap observables}.
            To this end, the symmetry of the measured histogram can be a useful rule of thumb as well as equilibration metrics, e.g., $\rho_t$ and others discussed in Refs.~\cite{machta:10a,wang:15a, weigel:21}.

        \subsubsection{Decay of bias with increasing \ensuremath{M}}\label{sec:num:bias:M}
    
            \begin{table}[b]
                    \centering
                    \caption{\label{tab:Bias_Mfit}
                    Exponents $b$ obtained from least-squares fits of the function $a \times M^{-b}$ to bias data in the Ising FM ($\beta \approx 0.44$) and SG ($\beta=3$), as explained in the main text.
                    Different equilibration levels are taken into account by varying the number of Metropolis sweeps $\theta$.
                    }
                    \begin{ruledtabular}
                    \begin{tabular}{cclccc}
                        \text{system} & L & \text{estimator} & $\theta$ & b & $\sigma(b)$\\
                        \colrule
                        FM & 64 & $\W{\widehat{e}}$         & 2     & 0.36 & 0.004 \\
                        FM & 64 & $\W{\widehat{e}}$         & 10    & 0.96 & 0.076 \\
                        FM & 64 & $\W{\widehat{f}}$         & 2     & 0.35 & 0.005\\
                        FM & 64 & $\W{\widehat{f}}$         & 10    & 1.02 & 0.099 \\
                        FM & 64 & $\Wvar{\widehat{c}}$      & 2     & 0.32 & 0.003 \\
                        FM & 64 & $\Wvar{\widehat{c}}$      & 10    & 0.96 & 0.060 \\
                        FM & 64 & $\Wvar{\widehat{\chi}}$   & 2     & 0.43 & 0.004 \\
                        FM & 64 & $\Wvar{\widehat{\chi}}$   & 10    & 0.96 & 0.061 \\
                        SG & 32 & $\W{\widehat{e}}$         & 2     & 0.53 & 0.076 \\
                        SG & 32 & $\W{\widehat{e}}$         & 5     & 0.79 & 0.047 \\
                        SG & 32 & $\W{\widehat{e}}$         & 10    & 0.88 & 0.026 \\
                        SG & 32 & $\W{\widehat{f}}$         & 2     & 0.60 & 0.020 \\
                        SG & 32 & $\W{\widehat{f}}$         & 5     & 0.87 & 0.029 \\
                        SG & 32 & $\W{\widehat{f}}$         & 10    & 0.75 & 0.023 \\
                        SG & 32 & $\Wvar{\widehat{c}}$      & 2     & 0.45 & 0.061 \\
                        SG & 32 & $\Wvar{\widehat{c}}$      & 5     & 0.76 & 0.108 \\
                        SG & 32 & $\Wvar{\widehat{c}}$      & 10    & 0.78 & 0.106 \\
                    \end{tabular}
                    \end{ruledtabular}
            \end{table}
            
            \begin{figure}
                \centering
                \includegraphics[width=\linewidth]{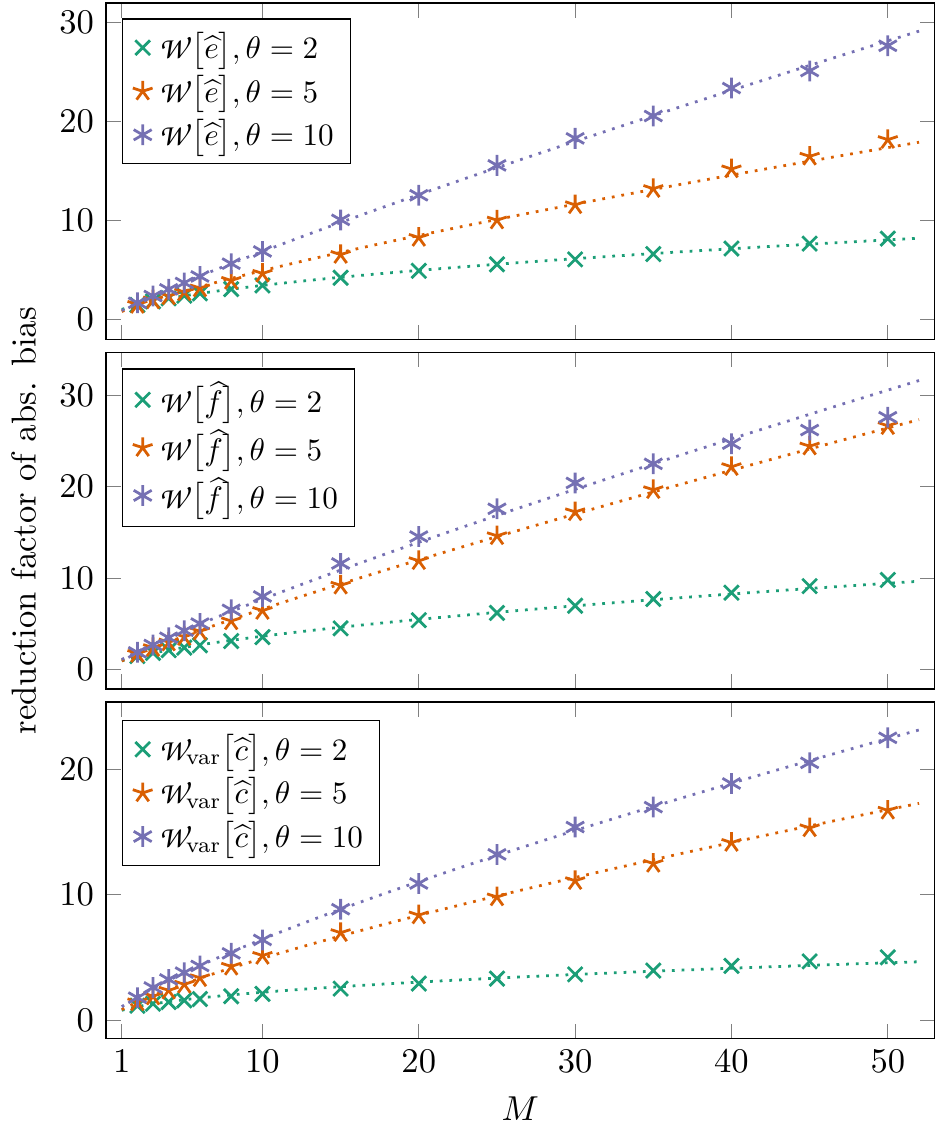}
                \caption{\label{fig:SG_Mscal}
                Ratio of the disorder averages of absolute systematic errors for single PA runs and for weighted averaging.
                50 randomly generated $L=32$ Ising SG instances at $\beta=3$ were simulated with the same number of $\theta$ Metropolis sweeps.
                Dotted lines represent the result of substituting the respective least-squares fit of $a \times M^{-b}$ (shown in Table~\ref{tab:Bias_Mfit}) into this ratio instead. 
                }
            \end{figure}
    
            Following this qualitative study, we quantitatively investigate the reduction of systematic errors with respect to the number $M$ of independent simulations over which the weighted average is performed.
            This is not only decisive for the efficiency of weighted averaging, but may also provide the appropriate value of $M$, if a certain bias level shall be reached.
            Wang \emph{et al.}~\cite{wang:15a} argued that systematic errors in configurational PA estimators as well as the free energy are proportional to $R^{-1}$ in the limit $R \to \infty$.
            They expected this relation to generalize to the respective weighted averages of $M$ runs with target population size $R$ by substituting $R \mapsto MR$~\cite{wang:15a}, indicating an asymptotic $M^{-1}$ dependence.
            
            The simplest way of testing these claims is to fix an inverse temperature $\beta_i$ and consider bias of a given observable at $\beta_i$ as a function of $M$.
            In fact, we closely follow this strategy for the Ising SG using the lowest temperature $\beta_f=3$ as bias is expected to be large in this regime.
            To prevent cancellation of systematic errors across different instances, we take the disorder average of the absolute values of bias measurements.
            Finally, we perform least-squares fits of the functional form $a \times M^{-b}$ to the bias data (see Fig.~\ref{fig:SG_Mscal} for a visual impression of the data and fits).
            The resulting exponents $b$ are shown in the lower part of Table~\ref{tab:Bias_Mfit} and the associated standard deviation $\sigma(b)$ was calculated by the jackknife~\cite{efron:82} approach applied to the set of 50 disorder realizations.
            As previously mentioned, the fact that choosing $R=2 \times 10^4$ and $\theta\leq 10$ is not sufficient for reliable $q$ measurements at low temperatures causes the bias reduction of weighted spin overlap estimators not to be monotonic with respect to $M$.
            We hence refrain from applying fits to the data for $q$.

            Regarding the Ising FM, systematic errors predominantly occur in the critical regime around $\beta \approx 0.44$ and can change sign, as demonstrated in Fig.~\ref{fig:c_chi_bias}.
            To get a notion of ``near-critical'' systematic errors, we decided to consider the absolute bias of a given observable averaged over the temperature range $0.42 \leq \beta \leq 0.46$.
            Due to the pronounced peaks of heat capacity and susceptibility near $\beta = 0.44$, this procedure was conducted for relative bias values.
            Statistical errors on this data were obtained by a bootstrapping approach and then entered the same procedure as described above for the Ising SG.
            The exponents estimating a proposed power-law decay of bias are shown in the upper half of Table~\ref{tab:Bias_Mfit} (in this case, $\sigma(b)$ relates to the standard fit error).
            
            First of all, correctly employed estimators for $e$, $f$, $c$ and $\chi$ always reduce systematic errors in the measurements we performed for Ising FM and SG --- in contrast to the results discussed for $P(q)$.
            Moreover, exponents obtained for the Ising FM seem to be relatively independent of the observable considered.
            However, the most crucial behavior displayed by both models is that \emph{the rate at which systematic errors are reduced by weighted averaging strongly depends on equilibration}.
            While measurements at $\theta=10$ often result in bias declining roughly proportional to $M^{-1}$, this relation must potentially be corrected to $1/\sqrt{M}$ or worse if simulations are far from equilibrium.
            This is consistent with the picture that free-energy weights are increasingly dominant in regimes with poor equilibration, which causes only few simulations to contribute to the weighted average~\cite{machta:10a,wang:15a,weigel:21}.
        
        \begin{figure}
                \centering
                \includegraphics[width=\linewidth]{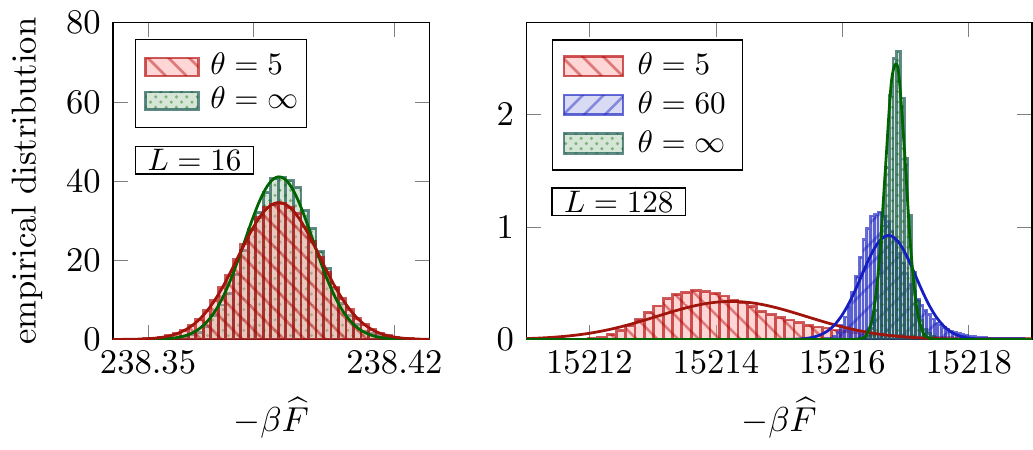}
                \caption{\label{fig:FM_bF_hist} Empirical distribution of $-\beta \widehat{F}$, i.e., the negative dimensionless free-energy for different system sizes of the Ising FM at $\beta=0.44$.
                Solid lines represent normal distributions with mean and variance given by the empirical distribution at the respective number of Metropolis sweeps $\theta$.}
        \end{figure}           
            
            When bias is proportional to $M^{-b}$ with $b \in (0,1)$, the ratio between error reduction and computational work worsens for larger $M$, which can be seen in Fig.~\ref{fig:SG_Mscal}.
            Herein, the reduction is calculated as the ratio of disorder-averaged absolute bias of single PA runs and weighted averaging at $\beta=3$.
            Dotted lines correspond to the least-squares-fits related to Table~\ref{tab:Bias_Mfit}.
            In this representation, the statistical error without sample-to-sample contributions is negligible, whereas the inclusion of such fluctuations causes certain error bars at $ \theta=10$ and $\theta=5$ to overlap.
            Thus, the data shown are very reliable for the fixed number of considered disorder realizations, whereas the number of 50 instances is hardly sufficient to generalize our results to a larger number of realizations.
            
            To better understand how free-energy weights work in the background, the distribution of $\widehat{F}$ and $w$ is studied in the next section.
            We want to emphasize that even for the moderate population size $R=2\times10^4$ we could not find any drawback with regards to bias from using the simplified free-energy weights $\underline{w}$ defined in Eq.~\eqref{eq:simpleWeight} instead of $w$.

    \subsection{Free energy and weight distribution}\label{sec:num:dist}
    
         Previous work predicts $\widehat{F}$ to be normally distributed if $R \to \infty$~\cite{wang:15a}.
        Such behavior is depicted for the Ising FM at $\beta=0.44$ in Fig.~\ref{fig:FM_bF_hist}.
        Since the state space of the small $L=16$ system is accurately sampled even if $\theta=5$ equilibration sweeps are performed at every temperature, the resulting empirical distribution of $-\beta \widehat{F}$ is remarkably close to the solid Gaussian curve having identical mean and variance.
        In contrast, the $L=128$ system in the right panel displays strongly skewed free-energy histograms at $\theta=5$ and $\theta=60$.
        Despite this poor sampling, weighted averaging reduces bias for $e$, $f$, $c$ and $\chi$ even for these parameters, albeit at a remarkably inefficient rate with respect to $M$ (not shown).
        
        \begin{figure}
                \centering
                \includegraphics[width=\linewidth]{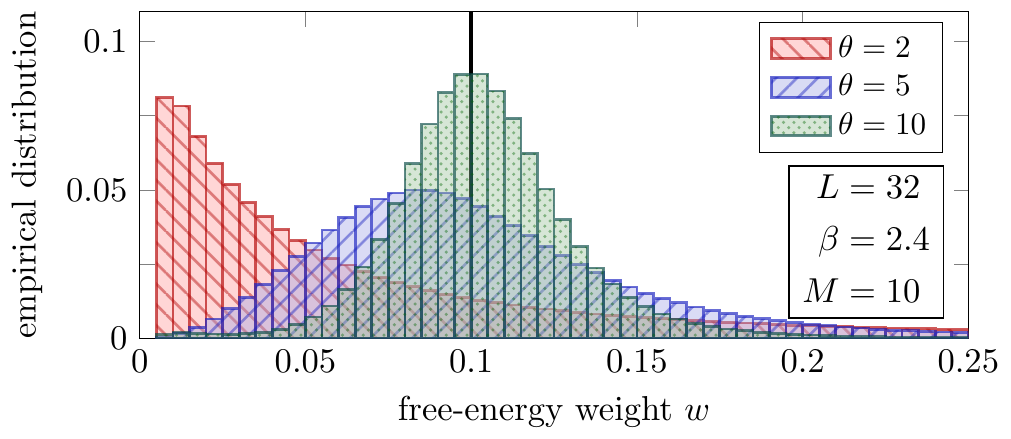}
                \caption{\label{fig:SG_whist} Disorder-averaged empirical distribution of free-energy weights for the SG model encountered during $S=5 \times 10^3$ weighted averages performed over $M=10$ runs.
                Every distribution consists of $2.5 \times 10^6$ individual PA runs and has a mean value of $1/M=0.1$ due to the normalization. 
                In the present histograms, a bin size of $200^{-1}$ is used.
                }
        \end{figure}
        
        A broad free-energy distribution has immediate consequences for the free-energy weights.
        If a PA simulation at the lower tail of the distribution was to be weighted against a counterpart from the upper tail, we may encounter weight ratios of $\exp(15 218-15 212)\approx 403$ in the right panel at $\theta=5$.
        In contrast, $\exp(238.42-238.35)\approx 1.07$ should be a reasonable upper bound for the $L=16$ system at the same number of Metropolis sweeps.
        Thus, we can confirm that equilibration and system size are crucial for the stability of weighted averaging~\cite{machta:10a, wang:15a, weigel:21}.
        
        Note that even in the limit $\theta \to \infty$, the finite value of $R$ results in a strictly positive variance of $\beta\widehat{F}$~\cite{weigel:21}, as demonstrated by the distributions filled with a dotted pattern in Fig.~\ref{fig:FM_bF_hist}.
        This $\theta=\infty$ limit is realized by replacing the Metropolis spin updates (used during step \ref{alg:step:iv} in Sec.~\ref{sec:ALG:alg}) by simple sampling of the energy density of states.
        This is possible since for the FM we have access to the exact energy distribution for finite systems~\cite{beale:96}.
        The remainder of the PA framework is unchanged and measurements are carried out in the same manner as for finite $\theta$.
        We refer to Ref.~\cite{gessert:22a} for a detailed discussion of this artificial setup.

        To see the effects of insufficient equilibration on the distribution of free-energy weights, consider Fig.~\ref{fig:SG_whist} showing weight histograms at different equilibration levels averaged over 50 instances of the Ising SG.
        Based on the relatively symmetric distribution at $\theta=10$, weight frequencies become increasingly skewed the further from equilibrium PA populations are.
        Most astonishingly, the histogram at $\theta=2$ displays a large tail and shows that the most probable weights are remarkably small. Note that, by construction, the weight histograms for each contributing realization have mean $1/M=0.1$. Hence such skewed shapes are indicative of individual contributing disorder realizations with similarly broad and skewed distributions.
        
        \begin{figure}
                \centering
                \includegraphics[width=\linewidth]{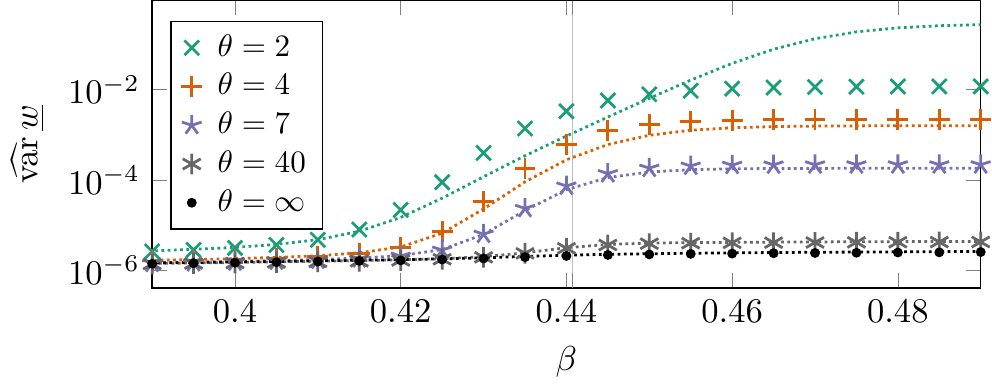}
                \caption{\label{fig:FM_varW} Sample variance of simplified free-energy weights $\underline{w}$ when averaging over $M=30$ runs in the critical regime of the $L=64$ Ising FM.
                Dashed lines represent the respective predictions of Eq.~\eqref{eq:varWpred} becoming increasingly inaccurate further from equilibrium.
                Simplified weights $\underline{w}$ were used as they are closer to the assumptions made in Sec.~\ref{sec:WA:w}.
                The difference between $w$ and $\underline{w}$ is marginal, however.
                The critical temperature $\beta_c \approx 0.4407$ is marked by a vertical line.                 
                }
        \end{figure}

        At least for smaller systems, the log-normality of free-energy weights can be conveniently checked by comparing the variance prediction formula Eq.~\eqref{eq:varWpred} to the actual variance.
        In doing so, data for the $L \in \{16, 32\}$ Ising FM are found to be in good agreement, whereas significant deviations start to occur at $L=64$, as is illustrated in Fig.~\ref{fig:FM_varW}.
        At large numbers of equilibration sweeps such as $\theta=40$, $\widehat{F}$ is normally distributed and our approximation is valid.
        While $\theta$ is lowered, however, more significant disagreements emerge spanning more than an order of magnitude at $\theta=2$ and $\beta\geq 0.48$.
        Similar behavior is observed at $L=128$.
        Generally speaking, Eq.~\eqref{eq:varWpred} is accurate if the distribution of $\widehat{F}$ is sufficiently narrow and close to Gaussian, i.e., for small systems, particularly well equilibrated runs or large population sizes.
        Therefore, it might still be helpful for simulations of the largest scale.
        
        The already mentioned unnoticeable difference with regards to bias reduction between using $\underline{w}$ and $w$ is reflected in mostly indistinguishable distributions (not shown).
        We attribute this to the small relative fluctuations of the population size that are observed already for the moderate value $R=2\times 10^4$ used here. 
        It is only in the limit $\theta \to \infty$ that the empirical distribution of $\underline{w}$ is slightly narrower than its non-simplified counterpart.
        This is plausible, since $w$ incorporates additional terms related to the population size which fluctuate independently of the exponential expression in Eq.~\eqref{eq:appropriateWeight}.

    \subsection{Statistical errors}\label{sec:num:stat}
    
        \begin{figure}
            \centering
            \includegraphics[width=\linewidth]{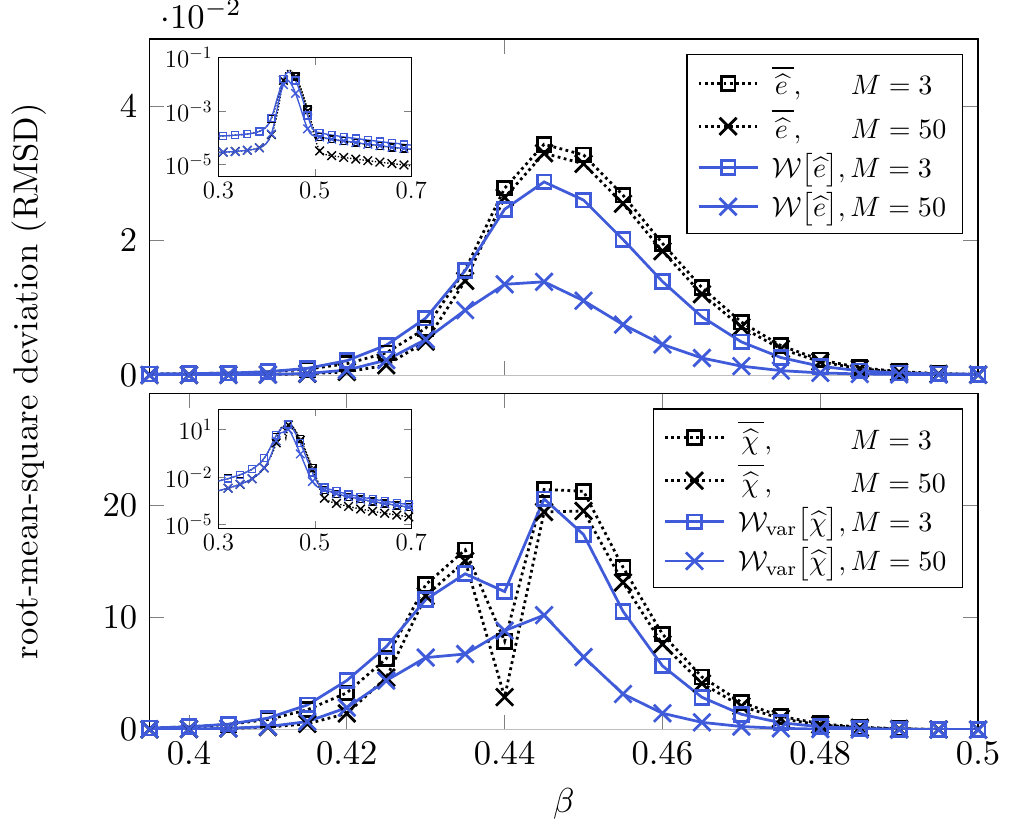}
            \caption{\label{fig:FM_RMSD}
            RMSD for different estimators of the internal energy (top) and susceptibility (bottom) in the $L=64$ Ising FM.
            Arithmetic averages (dashed lines) over $M$ independent runs employing $\theta=2$ equilibration sweeps are compared to the respective weighted estimators (solid lines). The insets show the same data on a logarithmic scale and a larger range of inverse temperatures.
            %In both insets, only every fifth data point is shown.
            }
        \end{figure}

        Besides the effect of diminished reduction rates of systematic errors, excessive fluctuations of the free-energy weights pose the threat of seriously increasing statistical errors, thereby potentially rendering weighted averaging practically useless~\cite{machta:10a,wang:15a,weigel:21}.
        In this section, we discuss to which extent these concerns are justified if populations in PA are far from equilibrium in simulations of the $L=64$ Ising FM and $L=32$ SG.
        
        To incorporate both systematic and statistical errors in one quantity, the \emph{root-mean-square deviation} is considered
        \begin{equation}\label{eq:RMSD}
            \text{RMSD}\coloneqq \sqrt{\text{bias}^2 + \text{variance}}.
        \end{equation}
        For both systems, we choose $\theta=2$ Metropolis sweeps at every temperature, which is largely insufficient for equilibration and results in dominant free-energy weights, as shown in Fig.~\ref{fig:SG_whist} at $\beta=2.4$.
        For the Ising FM, we mainly focus on the critical regime $\beta\approx0.44$, since systematic errors are very small everywhere else.
        
        \begin{figure}
            \centering
            \includegraphics[width=\linewidth]{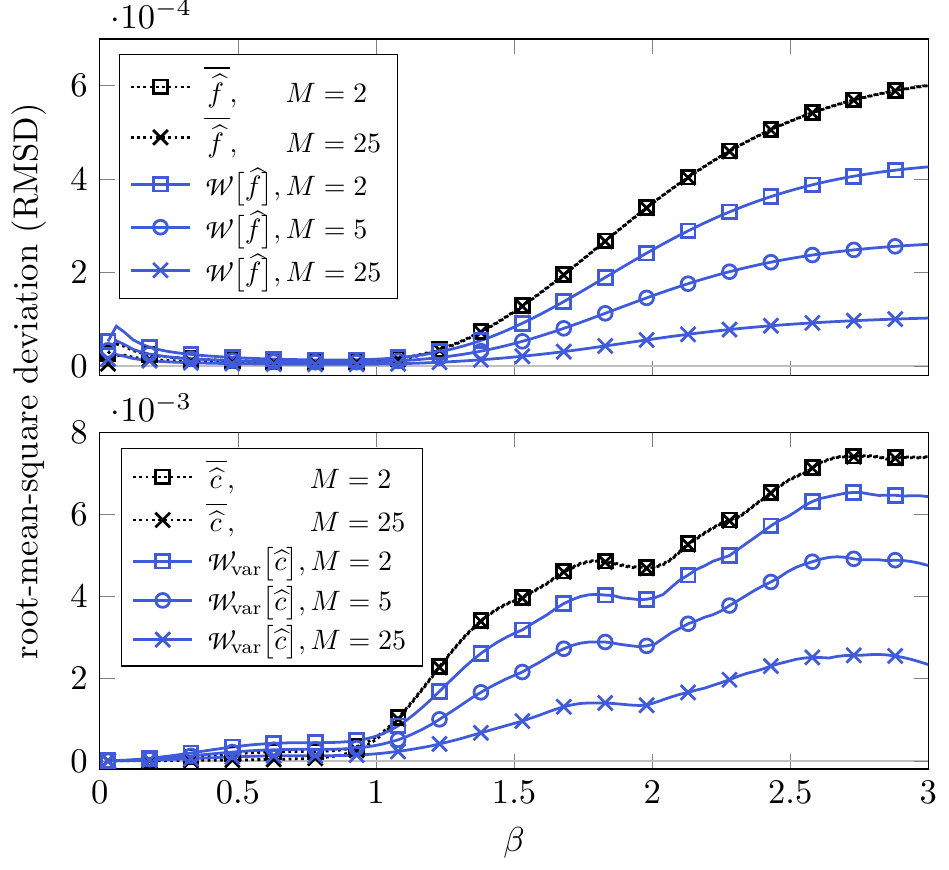}
            \caption{\label{fig:SG_RMSD}
            RMSD for different estimators of the free energy (top) and heat capacity (bottom) in the $L=32$ Ising SG.
            Arithmetic averaging (dashed lines) is compared to weighted averaging (solid lines) and only every fifth data point was drawn.
            At the present value of $\theta=2$, PA simulations do not properly sample the equilibrium distribution.
            A detailed description on the calculation of systematic and statistical errors is given in the main text.
            }
        \end{figure}       
        
        Fig.~\ref{fig:FM_RMSD} shows the RMSD increasing over three orders of magnitude as the annealing process approaches $\beta=\beta_c$.
        While there is no difference for $\beta \leq 0.4$, weighted energy and heat capacity estimators outperform arithmetic averages at most near-critical temperatures, even if only a small number of simulations is combined.
        Since arithmetic averages over $M=3$ and $M=50$ runs have similar RMSD values at the critical point, systematic errors dominate in this regime, demonstrating a clear advantage of weighted estimators.
        As a consequence of the incremental nature of $\widehat{F}$, weights remain dominant even at temperatures way below $\beta=0.5$.
        This results in the majority of simulations being effectively disregarded by the weighted average, while systematic errors are negligible even at $\theta=2$.
        Most drastically, the weighted average over $M=50$ runs behaves similarly to the arithmetic average over $M=3$ runs at $\beta \geq 0.5$ for both $e$ and $c$, as can be seen in the respective insets.
        Thus, one should always keep in mind that \emph{free-energy weights can remain exceedingly dominant after a regime of poor equilibration}, and one might want to reconsider whether weighted averaging should be employed at such temperatures.
        On the other hand, weighted estimators seem to provide better measurements in the critical regime, even if simulations are far from equilibrium.
        
        In case of the Ising SG, we decided to define bias in Eq.~\eqref{eq:RMSD} as the disorder average of the absolute value of systematic errors, similar to the consideration in Sec.~\ref{sec:num:bias:M}.
        This certainly results in larger bias values, but provides clearer evidence for the quality of measurements by preventing systematic errors for different realizations from canceling.
        The variance in Eq.~\eqref{eq:RMSD} is taken to be the sum of variances measured on every instance divided by the number of instances squared, i.e., no sample-to-sample contributions are taken into account, as we wish to solely consider this fixed set of realizations.
        
        Regarding the results of employing $\theta=2$ equilibration sweeps to all instances shown in Fig.~\ref{fig:SG_RMSD}, it is evident that systematic errors are the main source for deviations.
        This is consistent with the fact that bias for the ``hardest'' instances is usually larger than statistical fluctuations by a factor of 2 or 3 at these equilibration levels, i.e., it is not artificially created through our definition of bias.
        Consequently, weighted estimators outperform arithmetic averaging since the increased statistical error is overcompensated by the bias reduction. 
        On the other hand, we again observe that the gain-to-work ratio of weighted averaging with respect to $M$ is not particularly favorable, considering for instance the difference between $M=5$ and $M=25$.
        
        In summary, our numerical analysis shows that \emph{weighted averages can reliably outperform arithmetic averages even at poor equilibration levels, given that statistical errors are not larger than systematic deviations.}
        However, one should keep in mind ``memory effects'' of the free-energy weights as visible in Fig.~\ref{fig:SG_RMSD} when studying phase transitions.

\section{Conclusions}\label{sec:conclusion}

    We have provided an in-depth demonstration of the enhancement of population annealing measurements through weighted averaging.
    Since it only requires data which are already stored, namely the measured observable and the associated potential, the overhead of the method is marginal.
    Thus, just as population annealing itself, weighted averaging is highly compatible with massive parallelism and distributed systems.
    
    From a theoretical perspective, we established a rigorous mathematical foundation for weighted averaging and developed the notion of ``configurational'' estimators to emphasize that not every estimator can be weighted in the same manner to obtain asymptotically unbiased results.
    That is, not every weighted estimator is a weighted average of the corresponding estimators from individual PA simulations.
    Moreover, we rigorously proved that the method applies to a large family of target distributions in the setting of finite systems.
    For every observable considered so far, the appropriate weighted estimators could be derived by expressing the observable in terms of quantities whose weighted estimators are known (such as configurational estimators).
    This approach was demonstrated for central moments while we strongly suspect it to work for more involved quantities as well, e.g., the Binder parameter.
    In practice, Eq.~\eqref{eq:varWpred} might be helpful to predict the variance of free-energy weights in small systems or large populations, simultaneously allowing to probe log-normality of the weight distribution.
    
    Based on more than $10^7$ individual population annealing simulations of the two-dimensional Ising ferromagnet and spin glass we infer the following key observations:
    (i) Bias in energy, heat capacity, free energy and susceptibility measurements always decreased through appropriate weighted averaging.
    (ii) The method also worked for the spin overlap if population sizes were sufficiently large, but can even increase bias otherwise.
    Thus, we strongly recommend to carefully monitor the equilibration metrics discussed in Ref.~\cite{machta:10a,wang:15a, weigel:21} when performing weighted spin overlap measurements.
    (iii) Our data are in agreement with the picture~\cite{wang:15a} that systematic errors of correctly applied weighted estimators are roughly inversely proportional to the number of combined runs $M$ in well equilibrated settings.
    However, this dependency worsened far from equilibrium, potentially showing a closer resemblance to $1/\sqrt{M}$ or $1/\sqrt[3]{M}$.
    For reasonably equilibrated simulations and not too dominant free-energy weights, we even found that weighted averaging over $M$ runs can result in measurements practically indistinguishable from scaling the population size by $M$, as suggested in Ref.~\cite{wang:15a}.
    (iv) We could not find any drawbacks in using the simplified free-energy weights from Eq.~\eqref{eq:simpleWeight} when combining PA simulations of the same target population size, suggesting that prefactors related to the (fluctuating) population size do not matter; this is  consistent with the method used in Refs.~\cite{machta:10a, wang:15a, callaham:17, amey:18, rose:19}.
    (v) The feared breakdown of weighted averaging far from equilibrium was not observed, which is due to systematic errors dominating when both Ising systems are poorly equilibrated.
    Thus, even a mild bias reduction easily overcomes increasing statistical errors, thereby providing better estimates than the arithmetic average.
    Nevertheless, we expect this to change for very large systems or whenever dominant free-energy weights occur at times when statistical errors are prevalent.
    The latter case may happen if a regime of insufficient equilibration is followed by annealing steps where equilibration is easy, such as for the Ising ferromagnet.
    
    An additional approach to measure spin overlaps in single population annealing simulations was suggested and compared to ideas in Ref.~\cite{wang:15a}.
    Although it has larger statistical errors, it is easier, faster, parallelizable and compatible with the blocking analysis from Ref.~\cite{weigel:21}, rendering it superior for our use case.
    
    In conclusion, the extensive study of weighted averaging has proven once again the flexibility of population annealing as well as its potential for distributed computing.
    While parallelization allows trading hardware for time, weighted averaging enables the converse exchange if needed, such as compensating population sizes unachievable due to memory restrictions.
    Paired with plenty of room for different equilibration routines, annealing schedule tweaks and low-level optimization~\cite{barash:16,wang:15a,weigel:21,amey:18}, population annealing develops into an astonishingly fruitful simulation scheme suggesting that impressive applications lie ahead.

\begin{acknowledgments}
    We thank Lev Barash for providing us with the \texttt{PAising}-code and other program variants as well as Nico Heizmann for helpful discussions.
    Most calculations were performed using the Sulis Tier 2 HPC platform hosted by the Scientific Computing Research Technology Platform at the University of Warwick. Sulis is funded by EPSRC Grant EP/T022108/1 and the HPC Midlands+ consortium.
    Moreover, we acknowledge the provision of computing time on the parallel computer cluster \emph{Zeus} of Coventry University.
    The work of P.L.E. was supported by \emph{Gesellschaft der Freunde der TU Chemnitz}.
    D.G. acknowledges the support by the Deutsch-Französische Hochschule (DFH-UFA) through the Doctoral College ``$\mathbb{L}^4$'' under Grant No. CDFA-02-07. D.G. further acknowledges support by the Leipzig Graduate School of Natural Sciences ``BuildMoNa''.
\end{acknowledgments}

\appendix

\section{Free-energy bias for Ising ferromagnet}\label{app:F_bias}

    Consider the annealing schedule $0=\beta_0 < \beta_1 < \ldots$ applied to Ising ferromagnet on a lattice with (constant) coordination number $z$ and let $\Delta \beta_i \coloneqq \beta_i - \beta_{i-1}$.
    Then, we have $f \searrow -z/2$ for $\beta \to \infty$.
    If this is regarded in
    \begin{equation}\label{eq:f_asympt}
        - \beta_i \widehat{f}_i = \frac{1}{N}\ln Q_i - \beta_{i-1} \widehat{f}_{i-1},
    \end{equation}
    we obtain $\ln Q_i \approx (z/2)\Delta \beta_i  N$ asymptotically.
    Inserting this back into Eq.~\eqref{eq:f_asympt}, yields the asymptotic relation
    \begin{equation}
        \frac{\widehat{f}_i - \widehat{f}_{i-1}}{\Delta \beta_i} \approx -\frac{1}{\beta_i}(z/2+\widehat{f}_{i-1}).
    \end{equation}
    This is the discrete version of the differential equation
    \begin{equation}
        y'(x)=-\frac{z/2+y}{x},
    \end{equation}
    whose solutions are $y = C/x - z/2$.
    Hence, we obtain the asymptotic relation
    \begin{equation}
        \text{bias}\, \widehat{f}_i \approx \widehat{f}_i + z/2 \propto \beta^{-1}.    
    \end{equation}

\section{Calculations from Sec.~\ref{sec:WA:proof}}\label{app:proof}

    To shorten the notation, we omit the word ``fixed'' in the conditional expectation and denote by $J_k (\gamma)$ the set of indices of all replicas in $\gamma$ at $\beta_k$. 
    Note that $J_k (\gamma)$ has cardinality $R \widehat{\rho}_{k}(\gamma)$.
    Recall that we may assume $v_k(\gamma) > 0$ for all $k \leq i$ as explained in the main text.
    \begin{widetext}
    \textit{Deriving Eq.~\eqref{eq:proof:firstStep}}:
    Let $i \geq 1$ and recall that $Q_k$ is measured at $\beta_{k-1}$, i.e., if $\mathcal{P}_0, \ldots, \mathcal{P}_{i-1}$ are fixed, $Q_k$ is a constant for all $k \leq i$.
    \begin{subequations}
        \begin{align}
            \mathbb{E}\Big[ \widehat{\rho}_i (\gamma) \prod_{k=1}^i Q_k &\Big|\,\mathcal{P}_0, \ldots, \mathcal{P}_{i-1}\Big]
            = \mathbb{E}\Big[ \widehat{\rho}_i (\gamma) \Big|\,\mathcal{P}_0, \ldots, \mathcal{P}_{i-1}\Big] \prod_{k=1}^i Q_k\\
            &\overset{\text{\ref{item:proof:d}}}{=} \mathbb{E}\Big[ R ^{-1} \sum_{j \in J_{i-1} (\gamma)} r_i^{(j)} \Big|\,\mathcal{P}_0, \ldots, \mathcal{P}_{i-1}\Big] \prod_{k=1}^i Q_k
            = \Big(\sum_{j \in J_{i-1} (\gamma)} \mathbb{E}\Big[ r_i^{(j)} \Big|\,\mathcal{P}_0, \ldots, \mathcal{P}_{i-1}\Big] \Big)\, R ^{-1}\prod_{k=1}^i Q_k\\
            &\overset{\text{\ref{item:proof:c}}}{=}  \big(R \widehat{\rho}_{i-1} (\gamma)\big) \left(\frac{v_i(\gamma)/v_{i-1}(\gamma)}{Q_i}\right)\, R ^{-1} \prod_{k=1}^i Q_k
            = \frac{v_i(\gamma)}{v_{i-1}(\gamma)} \widehat{\rho}_{i-1}(\gamma) \prod_{k=1}^{i-1} Q_k. \label{eq:proof:AppB}
        \end{align}
    \end{subequations}
    \textit{Deriving Eq.~\eqref{eq:proof:secndStep}}:
    We denote the conditional probability of obtaining population $\mathcal{P}_{k}$ from the ancestor population $\mathcal{P}_{k-1}$ through resampling at the transition $\beta_{k-1} \mapsto \beta_k$ by $\mathbb{P}(\mathcal{P}_{k} | \mathcal{P}_0, \ldots, \mathcal{P}_{k-1})$ and define the set of reachable populations at $\beta_k$ for fixed ancestor populations
    \begin{equation*}
        \Gamma^R_k \coloneqq \{ \mathcal{P}_k \in \Gamma^R ~\vert~ \mathbb{P}(\mathcal{P}_{k} | \mathcal{P}_0, \ldots, \mathcal{P}_{k-1}) > 0 \}.
    \end{equation*}
    Now, using the law of total expectation in the first and third equality one obtains
    \begin{subequations}
        \begin{align}
            \mathbb{E} \Big[ \widehat{\rho}_i (\gamma) \prod_{k=1}^i Q_k \Big|\,\mathcal{P}_0, \ldots, \mathcal{P}_{i-2}\Big]
            &= \sum_{\mathcal{P}_{i-1}\in \Gamma_{i-1}^R} \mathbb{P}(\mathcal{P}_{i-1}|\mathcal{P}_0, \ldots,\mathcal{P}_{i-2}) \cdot \mathbb{E} \Big[ \widehat{\rho}_{i}(\gamma) \prod_{k=1}^{i} Q_k \Big| \,\mathcal{P}_0, \ldots, \mathcal{P}_{i-1}\Big]\\
            &\hspace*{-6pt}\overset{\eqref{eq:proof:AppB}}{=} \frac{v_i(\gamma)}{v_{i-1}(\gamma)}\sum_{\mathcal{P}_{i-1}\in \Gamma_{i-1}^R} \mathbb{P}(\mathcal{P}_{i-1}|\mathcal{P}_0, \ldots,\mathcal{P}_{i-2}) \cdot \Big( \widehat{\rho}_{i-1}(\gamma) \prod_{k=1}^{i-1} Q_k\Big)\\
            &= \frac{v_i(\gamma)}{v_{i-1}(\gamma)}~\mathbb{E} \Big[ \widehat{\rho}_{i-1} (\gamma) \prod_{k=1}^{i-1} Q_k \Big|\,\mathcal{P}_0, \ldots, \mathcal{P}_{i-2}\Big]\label{eq:proof:AppBB}.
        \end{align}
    \end{subequations}
    \textit{Recursion based on Eq.~\eqref{eq:proof:secndStep}}: Using the law of total expectation, we can shorten the sequence of fixed populations,
    \begin{subequations}
        \begin{align}
            \mathbb{E} \Big[ \widehat{\rho}_i (\gamma) \prod_{k=1}^i Q_k \Big|\,\mathcal{P}_0, \ldots, \mathcal{P}_{i-3}\Big]
            &= \sum_{\mathcal{P}_{i-2}\in \Gamma_{i-2}^R}  \mathbb{P}(\mathcal{P}_{i-2}|\mathcal{P}_0, \ldots,\mathcal{P}_{i-3}) \mathbb{E}\Big[ \widehat{\rho}_i (\gamma) \prod_{k=1}^i Q_k \Big|\,\mathcal{P}_0, \ldots, \mathcal{P}_{i-2}\Big]\\
            &\hspace*{-3.2pt}\overset{\eqref{eq:proof:AppBB}}{=} \sum_{\mathcal{P}_{i-2}\in \Gamma_{i-2}^R}  \mathbb{P}(\mathcal{P}_{i-2}|\mathcal{P}_0, \ldots,\mathcal{P}_{i-3}) \frac{v_i(\gamma)}{v_{i-1}(\gamma)} \mathbb{E}\Big[ \widehat{\rho}_{i-1} (\gamma) \prod_{k=1}^{i-1} Q_k \Big|\,\mathcal{P}_0, \ldots, \mathcal{P}_{i-2}\Big]\\
            &= \frac{v_i(\gamma)}{v_{i-1}(\gamma)}\mathbb{E}\Big[ \widehat{\rho}_{i-1} (\gamma) \prod_{k=1}^{i-1} Q_k \Big|\,\mathcal{P}_0, \ldots, \mathcal{P}_{i-3}\Big]\\
            &\hspace*{-3.2pt}\overset{\eqref{eq:proof:AppBB}}{=} \frac{v_i(\gamma)}{v_{i-1}(\gamma)}\frac{v_{i-1}(\gamma)}{v_{i-2}(\gamma)}\mathbb{E}\Big[ \widehat{\rho}_{i-2} (\gamma) \prod_{k=1}^{i-2} Q_k \Big|\,\mathcal{P}_0, \ldots, \mathcal{P}_{i-3}\Big],
        \end{align}
    \end{subequations}
    where the second invocation of \eqref{eq:proof:AppBB} substitutes $i$ by $i-1$. Repeat this until $\mathcal{P}_0$ is reached on the left hand side.

    \end{widetext}

% MAIN TEXT:

% REERENCES:
% https://doi.org/10.1021/acs.jctc.1c01198

\bibliography{paper}

\end{document}